\documentclass[gmd, manuscript]{copernicus}

\usepackage{rotating}  

\begin{document}

\nolinenumbers 

\title{AIMIP Phase 1: systematic evaluations of AI weather and climate models}

\Author[1][brianhenn@allenai.org]{Brian}{Henn} 
\Author[1]{Christopher S.}{Bretherton}
\Author[2]{Nikolay}{Koldunov}
\Author[3]{Christian}{Lessig}
\Author[4]{Maria J.}{Molina}
\Author[1]{Troy}{Arcomano}
\Author[1]{Oliver}{Watt-Meyer}
\Author[6]{Guillaume}{Couairon}
\Author[5,6]{Renu}{Singh}
\Author[7]{Robert}{Brunstein}
\Author[6]{Yana}{Hasson}
\Author[8]{Antonia}{Jost}
\Author[10]{Noah}{Brenowitz}
\Author[10]{Peter}{Manshausen}
\Author[9]{Nathaniel}{Cresswell-Clay}
\Author[9,10]{Dale}{Durran}
\Author[4]{Kyle Joseph}{Chen Hall}
\Author[11]{Janni}{Yuval}
\Author[11]{Dmitrii}{Kochkov}
\Author[11]{Stephan}{Hoyer}
\Author[11]{Ignacio}{Lopez-Gomez}

\affil[1]{Allen Institute for Artificial Intelligence, Seattle, WA, USA}
\affil[2]{Alfred Wegener Institute, Helmholtz Centre for Polar and Marine Research, Bremerhaven, Germany}
\affil[3]{European Center for Medium Range Weather Forecasts, Reading, UK}
\affil[4]{Department of Atmospheric and Oceanic Science, University of Maryland, College Park, MD, USA}
\affil[5]{INRIA, Paris, France}
\affil[6]{Google DeepMind, Paris, France}
\affil[7]{Otto-von-Guericke Universität, Magdeburg, Germany}
\affil[8]{University of Potsdam, Potsdam, Germany}
\affil[9]{University of Washington, Seattle, WA, USA}
\affil[10]{NVIDIA Corporation, Santa Clara, CA, USA}
\affil[11]{Google Research, Mountain View, CA, USA}




\runningtitle{TEXT}

\runningauthor{TEXT}

\received{}
\pubdiscuss{} 
\revised{}
\accepted{}
\published{}


\firstpage{1}

\maketitle

\begin{abstract}
We present the AI weather and climate model intercomparison project (AIMIP), phase 1. Drawing from the rich tradition of intercomparisons in climate model development, we specify a common experiment, output data format, and training constraints (namely, training against historical reanalysis data) for AIMIP Phase 1 models. We aim to identify differences in modeling frameworks and AI architectural choices that influence model behavior, and build trust in AI weather and climate models through open data and evaluation. AIMIP Phase 1 models must simulate the atmosphere given specified historical sea surface temperatures over 1979-2024. We evaluate the models' performance using five major evaluation criteria: biases, trends, response to El Ni\~{n}o-related sea surface temperature anomalies, temporal variability, and out-of-sample generalization tests. We find that the AI models are able to simulate the historical climate and response to forcing as well as a conventional physically-based model, but some AI models underestimate historical warming trends, and their predictions diverge in the out-of-sample generalization tests. We describe the AIMIP Phase 1 dataset that is publicly available for additional evaluations.
\end{abstract}


\introduction  

The Artificial Intelligence (AI) Model Intercomparison Project (MIP) was proposed to the broader scientific community in late 2024 by co-authors Koldunov and Lessig.  AIMIP aims to systematically evaluate and compare `AI weather and climate models' (AIWCMs), defined as AI-enabled models for predicting and projecting climate, including statistics and extremes of weather.  Several AIWCMs have been designed for simulating historical climate, e.g., for the purpose of seasonal forecasting; some groups are also targeting longer-term multidecadal climate-change projections. Thus, AIMIP is specifically focused on longer-term simulations than weather-focused evaluations \citep[e.g., WeatherBench and WP-MIP;][]{https://doi.org/10.1029/2023MS004019, mctaggartcowan2026wpmipartificialintelligencehybrid}. 

AIMIP uses analysis methodologies and output formats developed and accepted by the climate science community and especially Phase 7 of the Coupled Model Intercomparison Project (CMIP7), the flagship international project for evaluating and intercomparing physics-based climate models \citep{DunneCMIP7}.  We hope this will help steer the further development of AI-based climate models and socialize such models within the broader climate science community, achieving some of the goals laid out in \cite{Ullrich2025}.

This paper describes the protocol of the first phase of AIMIP and some initial results. AIMIP Phase 1 is an intercomparison in the style of the Atmospheric Model Intercomparison Project \citep[AMIP;][]{Gates1992, Gates1999} of AIWCMs trained on global reanalysis, and forced by historical patterns of sea surface temperature (SST) and sea ice concentration (SIC), during the satellite era of 1979-2024 \citep[and its revisions]{AMIP_forcing}. AMIP in 1989 was the first coordinated intercomparison of physically-based climate models; its design has proved enduringly useful for model evaluation and is now a standard part of ongoing CMIP intercomparisons. AIMIP Phase 1 leverages the existence of several AIWCMs that have recently documented capabilities for stable, accurate historical AMIP simulations, e.g., NeuralGCM \citep{Kochkov2024}, ACE2 \citep{Watt-Meyer2025}, DLESyM \citep{Cresswell-Clay2025}, and cBottle \citep{brenowitz2025climatebottlegenerativefoundation}.  

The AIMIP Phase 1 specification is deliberately patterned on the AMIP specifications in the CMIP DECK \citep{Eyring2016}, which is the set of benchmark simulations that must be performed for a climate model to be a part of CMIP.  This is envisioned as a first step toward AIWCMs that are interactively coupled to AI-based ocean and sea ice components across a range of likely future climates.  Such models could perform the full suite of DECK simulations required for participation in CMIP.   In addition to an AMIP simulation, the DECK includes four ocean-coupled simulations: pre-industrial; historically forced; 1\% per year CO$_2$ rise for 140 years from pre-industrial to fourfold larger concentrations; and abrupt CO$_2$ quadrupling.  Future phases of AIMIP may evaluate such coupled AI models once they are more mature. Another goal of AIMIP is to enable using analysis software designed within CMIP for evaluating physics-based climate models from standardized model outputs \citep[e.g., PMP and ESMValTool;][]{gmd-17-3919-2024, gmd-9-1747-2016}. 

Section \ref{sec:protocol} summarizes the AIMIP Phase 1 training protocol and simulations.  Section \ref{sec:models} introduces the AIWCMs that submitted results for this paper.  Section \ref{sec:results} compares some basic results about the accuracy of their climate means, trends, response to ENSO variability, and extrapolation to warmer climates, and Sect. \ref{sec:discussion} discusses the next steps needed for these AIWCMs to become useful tools for climate modeling.

\section{AIMIP Phase 1 goals and protocol}
\label{sec:protocol}

The specific goals of AIMIP Phase 1 are:

\begin{itemize}
    \item To systematically compare time-mean climate, climate trends, climate variability, and weather phenomena and extremes in multi-decadal `AMIP' simulations by AI-driven models of the global atmosphere trained exclusively on ERA5 reanalysis \citep{Hersbach2020} and forced by historical SST and SIC.
    \item To develop the capability of AIWCMs to output results in a common format compliant with CMIP7 variable names and conventions to enable comprehensive evaluation of AIWCMs by the broader climate science community.
    \item To enhance the visibility and credibility of AIWCMs and learn about their current strengths and weaknesses.
\end{itemize}

\subsection{What types of AIWCMs are in scope for AIMIP Phase 1?}
 
AIMIP Phase 1 does not restrict how AIWCMs are built, because AI models may have very different modeling approaches compared to conventional general circulation models (GCMs). These approaches may include:
\begin{itemize}
  \item Autoregressive full model replacement, which learns to time-march the entire three-dimensional atmospheric state 
  \item Hybrid architectures in which AI-based components interact with physics-based components, such as a dynamical core, that do not involve empirical tuning 
  \item Conditional sampling of a learned PDF of weather states from a time-evolving climate
\end{itemize}

From an AI perspective, training and evaluation of AIWCMs is a challenging problem. We are interested in the long-term statistics that make up the climate, but targeting them directly during training is computationally expensive, as they emerge from the aggregate of a very large number of snapshots of the atmospheric state. Thus, AIWCMs must develop strategies for making predictions of states that compose faithful representations of the climate and its boundary-forced evolution over long (decadal) timescales.

\subsection{AIMIP Phase 1 training and testing}

\subsubsection{ERA5 used for training and testing}
To make for meaningful comparison, the AI component of participating models must be trained exclusively on the ERA5 reanalysis from 1979 to 2014.  For consistency with this training protocol, all outputs are evaluated with respect to ERA5 as reference ground truth, acknowledging well-known imperfections of ERA5 for this purpose, such as that some variables \citep[e.g., surface precipitation,][]{Lavers2022} and some climate trends \citep[e.g.,][]{Loeb2022EvaluatingTTA, Allan2022GlobalCIA} are poorly constrained by observations.  

The main intercomparison analysis effort is focused on `standard’ AMIP-type model simulations that run continuously from late 1978 through 2024, which include a 3-month spinup period, the 36-year training period (1979-2014), and a 10-year out-of-sample test period (2015-2024). As described in Sect. \ref{sec:results}, for some metrics we compare model performance in the training and test periods, whereas in some cases we consider the entire period. 

\subsubsection{Monthly SST and SIC forcing datasets}
These simulations are forced using a custom-generated monthly-mean SST and SIC dataset \citep{arcomano_2025_17065758}.  This dataset is very similar to the CMIP7 obs4mips datasets for AMIP forcing \citep{DunneCMIP7} that ends in late 2022, but has been extended through 2024 and includes minor modifications that improve its behavior during seasonal transitions in SIC (see Appendix \ref{appendix:aimip_forcing}).

\subsubsection{Anthropogenic radiative forcing not used as an input feature}
Importantly, the standard simulations should not use the concentration of CO$_2$ or any other anthropogenic radiative forcer as an input feature during training, because of prior experience that this can lead to overfitting problems.  For instance, the steady rise of CO$_2$ during the training period can be co-opted by AI models as a proxy for a clock that allows it to learn the timing of individual events in the climate record, such as major El Ni\~nos.  This restriction (that SSTs and the seasonal cycle of insolation are the only historically varying forcings) follows the original AMIP protocol, but is no longer standard for physics-based atmospheric GCM simulations.  In the future, we hope to develop a model training protocol that allows this restriction to be relaxed.  Since global-mean SST has a trend that is correlated to the increasing net anthropogenic radiative forcing, it is still possible for the AIWCMs to learn effects of radiative forcing trends — such as changes in atmospheric state and top-of-atmosphere radiative fluxes — by leveraging their correlation with SST trends, regardless of whether there is a physical cause behind the correlations.

\subsubsection{Land surface simulation}

While physically-based models typically have an interactive land surface model in AMIP experiments, we do not require this for AIMIP Phase 1. Models may choose to simulate land-atmosphere interactions in any way so long as the required near-surface atmospheric variables (see Sec. \ref{aimip_required_outputs}) are provided.

\subsubsection{5-member ensemble}
Each model must generate a 5-member ensemble of the required simulation through whatever means they choose (e.g., initial condition perturbations, use of a stochastic model, model parameter perturbations, etc.) to assess unforced natural variability.  A larger ensemble may be necessary to robustly estimate a model's forced response and trajectory spread arising from climate variability \citep[as previously shown for conventional physics-based models;][]{deser2012uncertainty}, but the small AIMIP ensemble allows for an estimate of the dispersiveness of the AIWCMs. 

\subsubsection{$\Delta$SST = $+2 ~K$ and $+4 ~K$ simulations}
Following the `amip-p4k' experiments that are a part of the Cloud Feedback intercomparison \citep[CFMIP,][]{Webb2017} in CMIP, and to test the ability of AIWCMs to generalize to unseen conditions, we also request ensembles of AMIP-like simulations but with SST uniformly increased by $2 ~K$ and $4 ~K$. For AIWCMs trained purely on ERA5, these are out-of-sample tests, so large biases or instability are to be expected unless explicitly prevented by the model architecture; the models' performance in this context also may not be indicative of their utility in making predictions in the (near-) historical climate. There is no definitive ground truth for these simulations, but plausible climate changes from a uniform $+2 ~K$ or $+4 ~K$ SST can be estimated from physics-based climate models \citep{Sutton2007, Byrne2013}.

\subsubsection{Model documentation}
Models are expected to provide detailed documentation of their training approach, including hyperparameter tuning and model selection, as well as open weights for reproducibility.  Salient characteristics of models that submitted results in time to include in this paper are presented in Section \ref{sec:models}. 

\subsection{Requested outputs, format, and storage}
\label{aimip_required_outputs}

\subsubsection{CMIP compatibility}
AIMIP Phase 1 specifies detailed model output requirements to enable comparison and the use of sophisticated evaluation tools developed by the CMIP community for physics-based climate models. All submitting groups must provide data that generally follow CF-compliant \citep{eaton_2025_17801666} variable names and units, and CMIP-type file naming, variable grouping, and attribute conventions \citep{taylor_2025_15670624}, and are encouraged to use tooling such as implementations of the Climate Model Output Rewriter \citep[CMOR;][]{mauzey_2024_10946710} to do so.

AIWCMs predict a subset of the outputs produced by CMIP (in particular, many AIWCMs do not explicitly predict cloud properties), and this may affect their suitability for some general-purpose CMIP evaluation tools. The AIWCMs also use a diverse set of horizontal grids; following CMIP7 specification, the submission data are on each model's native horizontal grid, which may vary in its effective resolution. However, because we assess the geographical fidelity of simulated fields, it is required that horizontal resolution not exceed 500 km.  It is the responsibility of this evaluation to interpret and harmonize such outputs; see Sect. \ref{sec:evaluation_methods} for regridding methods.

\subsubsection{Three-dimensional fields}
Models are expected to provide outputs in each grid column at a set of CMIP-standard pressure levels, but may internally use a different set of variables.  The minimal set of pressure levels is 1000, 850, 700, 500, 250, 100, and 50 hPa, and the requested three-dimensional fields are temperature $T$, specific humidity $q$, and the eastward and northward wind components $u$ and $v$.  The 500 hPa geopotential height field is also a required output.  Other fields may also be output by individual models.

\subsubsection{Surface fields}
The following surface fields should be reported if the AIWCM predicts them:
\begin{itemize}
    \item surface pressure and/or sea-level pressure,
    \item surface temperature (skin temperature over land or sea ice, SST over ocean, and whatever the model uses in grid cells with mixed surface types),
    \item 2-meter air temperature, 
    \item 2-meter dewpoint temperature and/or 2-meter specific humidity,
    \item 10-meter eastward and northward wind components, and
    \item time-averaged surface precipitation rate, including both liquid and frozen.
\end{itemize}

\subsubsection{Output frequency}

For each ensemble member, we request:
\begin{itemize}
    \item monthly means of all requested outputs for the entire train and test period (Oct. 1978 through Dec. 2024), 
    \item daily means of these fields for 1 Oct. 1978 through 31 Dec. 1979, and for the year of 2024.  This daily output enables assessment of weather variability and extremes without dominating the overall storage requirements.
\end{itemize}

Monthly or daily time averages of these fields can be computed as the average of instantaneous samples from the AIWCMs at each prediction time, with the exception of time-averaged surface precipitation, which is defined as the average flux over the period.

\subsubsection{AIMIP Phase 1 dataset availability}
The complete submitted dataset is publicly available in an S3-compatible cloud store. This was generously enabled by DKRZ (the Deutsche Klimarechenzentrum in Hamburg, Germany). If there is sufficient interest, the data may also later be published to the Earth System Grid Federation \citep[ESGF,][]{CINQUINI2014400} for easier access by the CMIP community.

The storage requirements for the outputs submitted from each AI model depend on its horizontal and vertical grids.  A typical value can be estimated from the minimal requested outputs from each ensemble member of a standard simulation, namely four 7-level fields plus eight single-level fields written in single precision.  For a typical AIWCM with a 1\degree $\times$1\degree ~latitude-longitude grid, this request totals 5 GB for the monthly data and an additional 7.7 GB for the daily data. For a suite of 5 ensemble members, and additionally including ensembles of +2~K and +4~K simulations, the required storage would total 190 GB.  This will depend on the model grid resolution and structure, and any optional output fields.

\begin{sidewaystable}
\caption{AIMIP Phase 1 model submissions.}
\label{tab:submissions}

\scriptsize
\centering
\setlength{\tabcolsep}{3pt}

\begin{tabular}{p{2.2cm} p{2.0cm} p{2.8cm} p{1.6cm} p{1.6cm} p{2.3cm} p{3.0cm}}
\hline
Organization & Model name & References & Code & Temporal frequency$^{a}$ & Horizontal grid & Vertical grid \\
\hline

Ai2 &
ACE2.1-ERA5 &
\cite{Watt-Meyer2025} &
\cite{brian_henn_2026_19831256} &
daily, monthly &
$1^\circ \times 1^\circ$ &
13 pressure levels (‘gr’), regridded from native 8 model layers via ML; native layers also submitted (‘gn’) \\

ArchesWeather &
ArchesWeather &
\cite{Couairon2026, singh2026evaluatingskillstabilityarchesweather} &
\cite{guillaume_couairon_2026_20784771} &
daily, monthly &
$1^\circ \times 1^\circ$ &
7 pressure levels \\

ArchesWeather &
ArchesWeatherGen &
\cite{Couairon2026, singh2026evaluatingskillstabilityarchesweather} &
\cite{guillaume_couairon_2026_20784771} &
daily, monthly &
$1^\circ \times 1^\circ$ &
7 pressure levels \\

NVIDIA &
cBottle-1.3 &
\cite{brenowitz2025climatebottlegenerativefoundation} &
\cite{manshausen_2026_20832634} &
daily, monthly &
HEALPix order 6 ($\sim0.9^\circ$) &
8 pressure levels \\

University of Washington and NVIDIA &
DLESyM &
\cite{Cresswell-Clay2025} &
\cite{nathanielcresswellclay_2026_21270137} &
daily, monthly &
HEALPix order 6 ($\sim0.9^\circ$) &
Some surface, 850 hPa, and 500 hPa variables \\

University of Maryland (PARETO group) &
MD-1.5 v0.9 &
\cite{hall2026monthlydiffusionv09latent} &
\cite{kyle_hall_2026_21430292} &
monthly &
$1.5^\circ \times 1.5^\circ$ native; $1^\circ \times 1^\circ$ submitted (‘gr’) &
7 pressure levels \\

Google Research &
NeuralGCM &
\cite{Kochkov2024, Yuval2026} &
\cite{kochkov_2026_21303721} &
daily, monthly &
$2.8^\circ \times 2.8^\circ$ &
32 sigma levels native; 7 pressure levels submitted \\

Google Research &
NeuralGCM-HRD &
\cite{Kochkov2024, Yuval2026} &
\cite{kochkov_2026_21303721} &
daily, monthly &
$1^\circ \times 1^\circ$ &
32 sigma levels native; 7 pressure levels submitted \\

NOAA GFDL &
CM4 (AM4 component) &
\cite{https://doi.org/10.1002/2017MS001208, https://doi.org/10.1029/2019MS001829} &
N/A$^{b}$ &
daily, monthly &
C96 cubed sphere native; $\sim1^\circ \times \sim1^\circ$ evaluated here &
33 sigma levels native; 7 pressure levels evaluated here \\

\hline
\end{tabular}

\vspace{0.5em}

\raggedright
\footnotesize
$^{a}$ Monthly indicates monthly average output from 1 Oct. 1978 to 31 Dec. 2024; daily indicates daily average output over (1) 1 Oct. 1978 to 31 Dec. 1979 and (2) 1 Jan. 2024 to 31 Dec. 2024. 

$^{b}$ Previously published CMIP6 output used; see Code and data availability.

\end{sidewaystable}

\section{Participating Models}
\label{sec:models}

AIMIP Phase 1 received submissions for eight AIWCMs from six model development groups. Table \ref{tab:submissions} summarizes the AIMIP Phase 1 submissions, which are also described in more detail in the remainder of this section.

\subsection{ACE2.1-ERA5}

ACE2.1-ERA5 is an autoregressive, SFNO-based \citep{pmlr-v202-bonev23a} emulator with a 6-hourly timestep. It is a variant of the ACE2-ERA5 model described in \cite{Watt-Meyer2025}, with the following differences in training methods:
\begin{itemize}
    \item CO$_2$ is not included as a forcing feature for ML predictions.
    \item The training period is adjusted to reflect the AIMIP specification: training is conducted over timesteps starting from Jan. 1 1979 and ending on Dec. 31 2008; validation and inline inference are conducted from Jan. 1 2009 to Dec. 31 2014.  
    \item The deterministic SFNO architecture uses layer normalization between blocks instead of instance normalization.
    \item Near-surface variables (2-meter air temperature and specific humidity and 10-meter winds) that were prognostic in ACE2-ERA5 are not prognostic, but instead learned as secondary diagnostics (see next point).
    \item Additional diagnostic variables that were not on ACE2's native model layer grid were predicted via a learned secondary decoder, a small-capacity gridpoint-local MLP without feedback upon the primary network weights, from the primary prognostic and diagnostic outputs. The additional predicted diagnostics include the near-surface variables listed above, as well as temperature, humidity, and winds at each required pressure level, and 500 hPa geopotential height.
\end{itemize}
The AIMIP SST and SIC forcing were regridded to ACE2.1-ERA5's native 1\degree ~Gaussian grid (via conservative regridding) and 6-hourly timestep (via linear interpolation). For each experiment, 5 ensemble members were generated by lagging ERA5 initial conditions from the 5-day window centered on the AIMIP start date of Oct. 1 1978. As in \cite{Watt-Meyer2025}, ACE2.1-ERA5 applies SST forcing by overwriting the prognostic surface temperature variable with the prescribed SST values for grid cells that are majority ocean.

\subsection{ArchesWeather and ArchesWeatherGen}

ArchesWeather is a data-driven deterministic weather forecasting model, and ArchesWeatherGen is a probabilistic model that leverages the deterministic forecasts of ArchesWeather \citep{Couairon2026}. Both models use the same neural network architecture consisting of a convolutional encoder-decoder and a SwinTransformer with earth-specific attention mechanism. ArchesWeather is trained with a mean-squared error loss to predict the state of the atmosphere $x_{t+1}$ at time $t+1$ given initial conditions $x_{t-1}$ and $x_{t}$ and a time conditioning $t_{cond}$. The model is then used auto-regressively to generate trajectories for the future weather. For ArchesWeatherGen, a flow matching-based generative module is trained to sample state residuals, i.e., $r_{t} = \hat{x}_{t} - x_{t}$ where $\hat{x}_{t}$ is the prediction of an ArchesWeather model (or the ensemble mean of such models).  This residual approach employed for ArchesWeatherGen enables faster training compared to a full-state generative model and enables ensemble predictions due to its probabilistic nature.

For AIMIP, the following changes were made to the ArchesWeather and ArchesWeatherGen pipeline \citep{singh2026evaluatingskillstabilityarchesweather}:
\begin{itemize}
    \item Models are trained on daily average ERA5 data instead of instantaneous states. To maintain the same amount of training data, our daily averaged data is calculated with a moving window approach containing four states (e.g., z00, z06, z12, z18).
    \item Models are trained on the 1\degree x1\degree ~native grid resolution used for evaluation within AIMIP. We increased the kernel size and stride of the convolutional embedding layer from 2 to 3, to ensure that the models operate on the same number of tokens as the default 1.5\degree ~version.
    \item SST and SIC were added as additional prognostic variables. Therefore, the model is informed not only by the monthly mean forcings prescribed by AIMIP but also by the recent state of the sea surface.
    \item NaN values in the state (e.g., SST over land and sea ice) are replaced by climatological means over time and space. In the perturbed SST experiments these climatological fill values are not perturbed. In the loss computation, these values are masked, i.e., the model is not trained to predict values where there are NaNs in the input state.
    \item Finally, the original month time conditioning was changed to day of year, which gives more information to the model.
    
\end{itemize}

For ArchesWeatherGen, its probabilistic nature was used to generate the different ensemble members by altering the seed with which the noise is sampled. For ArchesWeather, the members were simulated by selecting the dates from the 29th of September 1978 until the 3rd of October and used these as initial conditions.

\subsection{cBottle-1.3}
\label{cBottle}

cBottle is a diffusion model that predicts snapshots of the state fields at varying resolutions conditional on SST fields and other inputs \citep{brenowitz2025climatebottlegenerativefoundation}; cBottle-1.3 is a variant that meets the AIMIP specification. As in the original cBottle, cBottle-1.3 is generated on a HEALPix \citep{Gorski_2005} grid of order 6 ($n_{side}=64$), or an approximate equivalent resolution of 0.9\degree. For the AIMIP submission, there are eight snapshots per day; surface precipitation is trained on the ERA5 hourly accumulated precipitation and converted to units of mass flux.  

\textit{Training}: The model was retrained with a new train-test split respecting the AIMIP specification (i.e., split after 2014). Further, specific humidity at pressure levels, 2-meter dewpoint temperature, surface pressure, and skin (surface) temperature were added. Outputs are on the required pressure levels for AIMIP, except for the 250 and 100 hPa levels; for the evaluations here these were interpolated from the submitted 300, 200, and 50 hPa levels. Model channels were increased from 196 to 256. The AIMIP SST and SIC dataset was used as conditioning in this training, rather than the input4MIPs AMIP fields previously used in training. The SST values over land were filled with a constant 290 K value, and were not changed in the perturbed SST experiments. 

\textit{Inference}: cBottle-1.3 includes five ensemble members, created from different combinations of training checkpoints and using the native `uncorrelated', or the new `correlated' inference. The checkpoint and inference configuration are mapped onto the `physics\_index' in the AIMIP outputs as the model ensemble. Unlike autoregressive AI models, cBottle generates independent samples at each inference. This means that the resulting monthly means are expected to have much lower variance than an autoregressive model would. Therefore, a technique is added here to reintroduce temporal correlation at inference time. Physics indices 1-4 are run in `correlated noise' inference mode, where the random latent used in diffusion model sampling is correlated in time. This is a notable change with respect to the results in \cite{brenowitz2025climatebottlegenerativefoundation}. The latents are generated according to an autoregressive process of order one (AR1), where the current latent is a linear combination of the previous one and a random noise perturbation. This is scaled such that the overall variance is constant in time. More specifically:
\begin{gather*}
    x_t = \phi * x_{t-1} + sqrt(1-{\phi}^2) * {\epsilon}_t \quad \text{and} \quad \phi = 2^{(-1/\lambda)}
\end{gather*}
with latents $x_t$, correlation strength $\phi$, Gaussian noise $\epsilon$, and half-life $\lambda$. The same correlated latents are used for physics indices 1-4. We perform the inference at three-hourly frequency and choose the half-life to be 8 steps, i.e., 24 h. This means the autocorrelation of the latents drops by half in 24 h. For comparison, physics index 5 is run in native uncorrelated mode, where each latent is independent. While this inference-time temporal correlation produces smoothly varying atmospheric states, their temporal correlation is not expected to bring about correct atmospheric dynamics; this would require autoregressive training. When variability of temporal means are presented here, note that these will depend on our choice of the correlation half life, with a larger half-lifer corresponding to more variability of the temporal means. See the Appendix \ref{appendix:cBottle_indices} for a list of the checkpoints used in each physics index.

\subsection{DLESyM}

The Deep Learning Earth System Model \citep[DLESyM,][]{Cresswell-Clay2025} uses coupled convolutional neural networks built with ConvNeXt blocks \citep{liu2022convnet2020s, Karlbauer2024} and gated recurrent units. Model layers are organized in a U-Net framework. Atmospheric forecasts are generated autoregressively with 6-hourly time resolution. The model is trained to minimize RMSE loss over one 24-hour period.  The five-member ensemble is generated using initial conditions from five successive days in October 1978.  DLESyM was trained using outgoing long-wave radiation (OLR) from a satellite record starting in 1983. Here, the initial OLR fields were specified using ERA5's top net thermal radiation (TTR) field. 

DLESyM is designed for simulation of the earth system and is a coupled atmosphere-ocean model. In simulations for AIMIP, the ocean component is replaced by the prescribed SST field. Monthly SST forcing data were linearly interpolated to 96-hour resolution for compatibility with the coupled time stepping scheme during inference. During training and inference, SSTs over the continent are filled by zonal-linear interpolation and passed to the ocean module to provide values for  convolutional operators near the coasts. These same interpolated values, together with the rest of the SST field, are also passed to the atmospheric module ensuring fully defined convolutional operations over the globe. A land-sea mask encourages, but does not force, the atmospheric module to ignore the infilled SSTs over land. Those SSTs are constant over the 24-hour atmospheric loss-function calculation during training, and over a 96-hour period during inference. In the perturbed SST experiments the perturbation is applied on this filled field. Due to its intentionally parsimonious set of predicted fields, DLESyM is not directly able to produce the full set of variables specified by AIMIP. Here, we evaluate only near-surface temperature, 850 hPa temperature, 500 hPa geopotential height, and surface precipitation. Like cBottle1.3, DLESyM is generated on a HEALPix grid with an approximate equivalent resolution of 0.9\degree.    

The exact checkpoint of DLESyM described in \citet{Cresswell-Clay2025} is used here, which means that DLESyM is trained on data from Jan. 1 1983 through Jun. 30 2016. This training dataset includes 1.5 of the 10 years of AIMIP holdout data (which begins in 2015). In this respect it does not conform to the AIMIP specification in the way that the other models do, and thus may gain some relative skill in the holdout evaluations, but on the other hand DLESyM does not use the 1979-1982 period for training.

\subsection{Monthly Diffusion at 1.5\degree ~resolution (MD-1.5 v0.9)}

Monthly Diffusion at 1.5\degree ~resolution \citep[MD-1.5 v0.9,][]{hall2026monthlydiffusionv09latent} is an autoregressive latent diffusion model developed for AIMIP Phase 1. The model advances the atmospheric state at a one-month timestep and is designed to emulate low-frequency atmospheric variability in a computationally efficient framework. Rather than performing autoregression directly in physical space, MD-1.5 v0.9 operates in a learned latent space. The atmospheric state is first encoded into a compact latent representation, which is then advanced by one month using a conditional denoising network, and the predicted latent is decoded back into physical variables at each desired sampling time.

MD-1.5 v0.9 is conditioned on both external forcings and slowly varying boundary information. Specifically, MD-1.5 v0.9 uses the AIMIP SST and SIC forcings along with learned seasonality embeddings. It also incorporates a land-sea mask and time-invariant topographical fields as conditioning tensors within the architecture. These conditioning tensors are used during encoding, autoregressive latent prediction, and decoding, allowing the model to represent atmospheric variability that depends on both prescribed oceanic forcing and the seasonal cycle. This architectural choice may reduce the extent to which the latent representation must capture seasonally recurring and ocean-forced structure on its own, potentially allowing it to focus more on internally generated atmospheric variability.

The MD-1.5 v0.9 training procedure fills missing values in the SST forcing with a constant value of zero (climatological mean over time and space) after normalization; these filled values are not changed in the perturbed SST experiments. It then provides the model with an explicit binary validity mask, where a value of one indicates a valid cell and a value of zero indicates a missing value. The binary mask is concatenated with SST and SIC as an additional forcing channel, to enable the model to learn proper treatment of missing values. 

Architecturally, MD-1.5 v0.9 consists of three neural networks: an encoder, a decoder, and a denoising predictor. The encoder and decoder form a conditional variational autoencoder, while the predictor is a conditional latent diffusion model that evolves the latent state forward in time. All three networks use low-rank spectral operators based on SFNO, making the model suited to global gridded data on the sphere.

Unlike many latent diffusion models \citep{rombach2022highresolutionimagesynthesislatent}, which are trained in two stages by first fitting an autoencoder and then fitting a diffusion model on the resulting latent space, MD-1.5 v0.9 is trained jointly end-to-end. In this data-limited monthly-mean setting, joint optimization encourages the encoder to learn not only a compact representation of the atmospheric state, but also a latent geometry that supports smooth conditional diffusion dynamics. The denoising network also learns to center and scale the latents online so that they are compatible with independent and identically distributed DDPM (Denoising Diffusion Probabilistic Model) noise, rather than relying on offline latent statistics from a separately pretrained autoencoder. Since the latent representation continues to evolve during training, we maintain an exponential moving average of the model weights to improve robustness.

The model’s prognostic state consists of monthly-mean values of seven surface-level variables (skin temperature, surface precipitation, 2-meter air and dewpoint temperatures, surface pressure, and 10-meter $u$ and $v$ winds) and five atmospheric variables (temperature, specific humidity, geopotential height, and wind components) on the seven AIMIP protocol-specified pressure levels. Training data were derived from ERA5 monthly means on a 1.5-degree equiangular grid. Contiguous training and validation splits were used, with the training period from Jan. 1 1985 through Dec. 31 2014 and the validation period from Jan. 1 1979 through Dec. 31 1984.

Daily data are not provided for MD-1.5 v0.9.

\subsection{NeuralGCM and NeuralGCM-HRD}

The submitted models build upon previous versions of NeuralGCM described in \citet{Kochkov2024} and \citet{Yuval2026}. Their main novelty is the addition of machine-learned output heads that enable the decoding of diagnostic variables from the prognostic atmospheric state, which operates on a native 2.8\degree ~regular grid in both NeuralGCM and NeuralGCM-HRD.

The architecture of the decoder heads follows the learned physics module used to predict tendencies in \cite{Kochkov2024}, but exclude the convolutional-based embeddings and use independent Gaussian Random Fields (GRFs) to model stochasticity in observations. The output heads consume the same features (prognostic fields, radiation, etc.) as the learned physics module, and are optimized jointly with it using a Continuous Ranked Probability Score (CRPS) loss. 

Two model variants were used to generate AIMIP Phase 1 submissions. The \textit{NeuralGCM} variant outputs all fields on the model's native 2.8\degree ~regular grid, and leverages decoder heads to infer the mean evaporation rate, 2-meter temperature, surface pressure, skin (surface) temperature, 2-meter dewpoint temperature, and 10-meter $u$ and $v$ wind components. The \textit{NeuralGCM-HRD} variant downscales both surface and pressure-level variables to a resolution of 1\degree. The high-resolution decoder head (HRD) takes the full, native 3D prognostic atmospheric state operating on the model's sigma levels as input to the neural decoder. These are regridded to 1\degree ~resolution. The upsampled prognostic fields are combined with high-resolution auxiliary inputs including fine-scale orography, latitudinal coordinates, high-resolution learned surface feature embeddings, and an independent Gaussian Random Field (GRF) and used as inputs. A scale correction is applied such that the high-resolution fields of 2-meter temperature and precipitation, when coarse-grained, strictly match the corresponding outputs generated by the low-resolution heads (in the future this can be extended to other fields). Note that a bug in the code prevented the random field for this specific head from updating across time steps (freezing the noise) during training, but the GRFs were updated correctly during inference.

Precipitation rates are diagnosed to ensure consistency with the total column water budget, with small deviations from the method described in \cite{Yuval2026}. Specifically, precipitation ($P$) is derived from the column water budget ($P - E$, computed by the dynamical core) and the predicted evaporation ($E$, from the surface head). In addition, a strict non-negativity constraint ($P \geq 0$) is applied to the result. If the diagnosed precipitation is negative, the model dynamically clips the predicted evaporation field to close the budget. Finally, the precipitation rate is trained against ERA5 precipitation to ensure consistency with the AIMIP protocol.

In AIMIP, the prescribed SST values are merged with land and sea ice embeddings as described in \cite{Kochkov2024}, such that no missing values appear in the merged embeddings (i.e., no fill values are required).

\subsection{CMIP6 model: GFDL-CM4}

We include the output of one physically-based global atmospheric model for comparison, the NOAA GFDL CM4 \citep{https://doi.org/10.1029/2019MS001829} model's submission to AMIP in CMIP6 \citep{https://doi.org/10.22033/ESGF/CMIP6.8494}. Specifically, we use the AM4 atmospheric model \citep{https://doi.org/10.1002/2017MS001208} output from the CM4 AMIP and perturbed SST simulations, as these have more complete output fields than the AM4-labeled AMIP simulations, though both are prescribed SST atmospheric simulations. AM4 in CM4 was run at approximately 1\degree ~resolution (cubed sphere C96), similar to the horizontal resolution of most of the AIWCMs here, and with 33 vertical levels. Unlike the 5-member AIWCM ensembles, one GFDL-CM4 AMIP ensemble member is available.

\section{AIMIP Phase 1 Evaluation Results}
\label{sec:results}

\subsection{Evaluation metrics and methods}
\label{sec:evaluation_methods}

The following evaluations were conducted as part of AIMIP Phase 1:
\begin{itemize}
    \item{E1:} Train-period and test-period time-mean bias patterns and global area-weighted bias pattern RMS (`root mean square bias' or RMSB), vs. ERA5 reference.
    \item{E2:} Train-period and test-period linear trends. Trends are computed based on series of global area-weighted annual mean values, and maps of gridpoint trends are also computed. 
    \item{E3:} Regressions of predicted fields on the Ni\~{n}o3.4 index \citep{Barnston1997} computed from the specified AIMIP SSTs.
    \item{E4:} Maps and global means of temporal standard deviation of selected variables, calculated using differences of the daily data from the monthly mean data. 
    \item{E5:} Atmospheric response to $+2 ~K$ and $+4 ~K$ uniform SST perturbation experiments.
\end{itemize}

Items E1, E2, E3, and E5 are computed using the 1979-2024 monthly-mean predictions. Item E4 uses the 1979 daily-mean predictions as well.

Because of the disparate grids upon which the AIWCMs submit data, a regridding process is necessary in order to compute differences between the models and ERA5 or between models. This process was conducted by first conservatively regridding (coarsening) the relevant ERA5 fields from the regular 0.25\degree ~grid to a regular 1\degree ~grid, and also coarsening the hourly ERA5 data to monthly or daily resolution as appropriate. For metric evaluation, relevant quantities (time-means, trends, etc.) were first computed on each model or dataset's native grid. Then, for computation of error metrics, each model's data were regridded onto the ERA5 grid via conservative regridding using the xESMF package \citep{zhuang_xesmf_2020}, except for HEALPix gridded models, where the regridding was not conservative but nearest-neighbor. 

While most of the models' native grids are at approximately 1\degree ~resolution, including NeuralGCM-HRD, NeuralGCM produces outputs at 2.8\degree ~resolution, and shares its underlying prognostic state with NeuralGCM-HRD. Therefore, we consider 1\degree ~resolution NeuralGCM-HRD in the primary metrics presented here. However, NeuralGCM results are shown in Appendix \ref{app_2p8deg_results}; in that case ERA5 and the other models' data were first regridded to NeuralGCM's 2.8\degree ~resolution grid for computation. Regardless, in most cases the resulting metrics are not strongly resolution-dependent. 

Additionally, because CMIP6 CM4 variables on pressure levels are not filled with extrapolated values when the levels are below the ground surface (unlike ERA5, which does extrapolate), we mask the analysis of pressure-level variables for all models to times and places where it is above the surface in CM4. 

\subsection{E1: Biases}

\begin{figure*}[h]
\includegraphics[width=17.4cm]{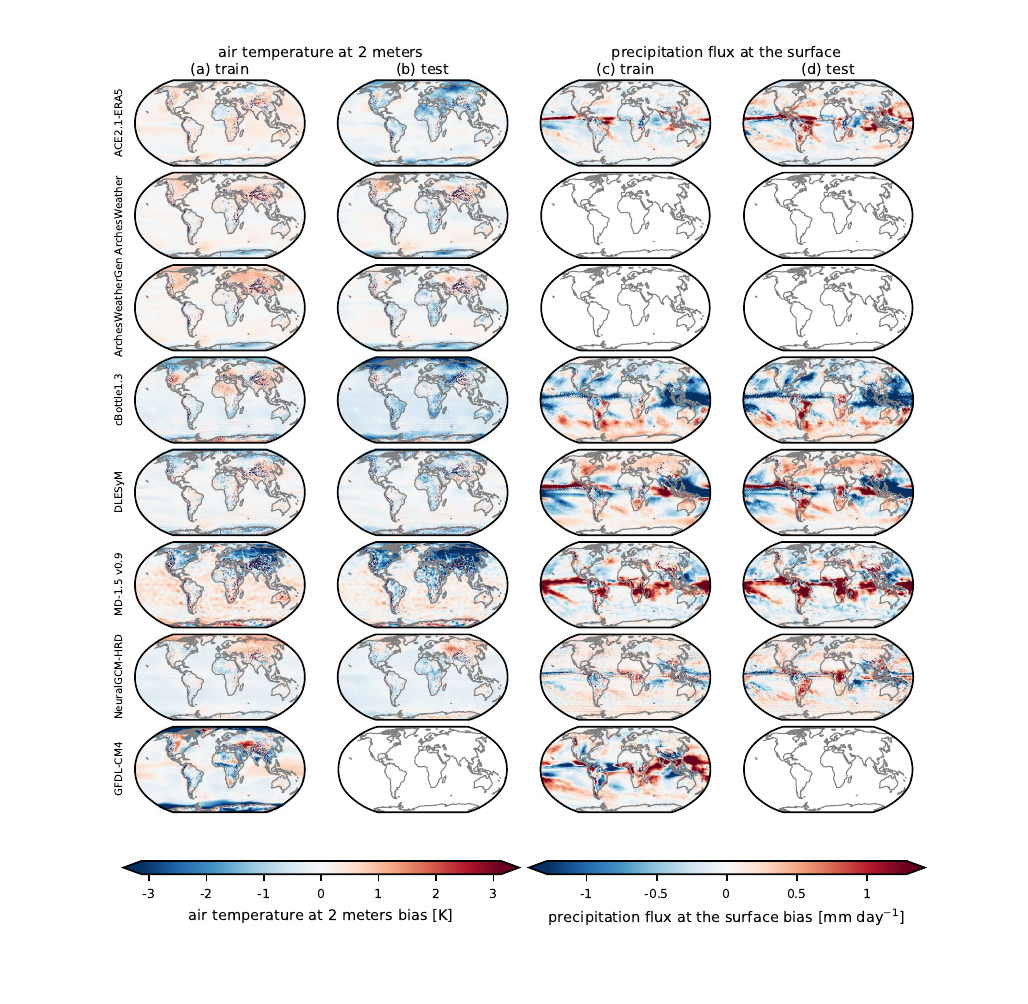}
\caption{Biases at 1\degree ~resolution versus ERA5, for the AIWCMs and a CMIP6 model (GFDL-CM4, bottom row). (a), (b): 2-meter air temperature biases over the training (1979-2014) and test (2015-2024) periods, respectively. GFDL-CM4 data end in 2014 and so are only available over the training period. (c), (d): surface precipitation biases over the same periods, for models that included surface precipitation outputs (ArchesWeather and ArchesWeatherGen did not).}
\label{fig:bias_map}
\end{figure*}

\begin{figure*}[h]
\includegraphics[width=17.4cm]{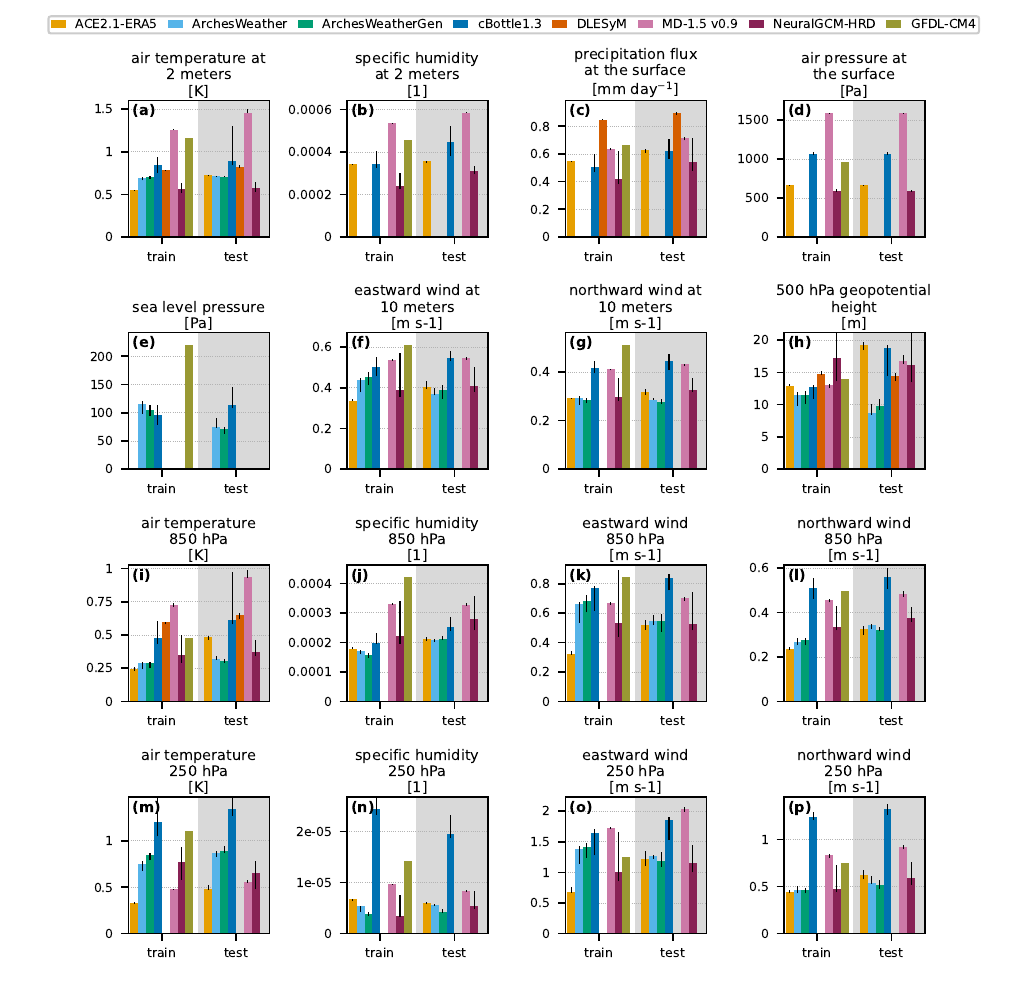}
\caption{RMSB area-weighted over the globe on the 1\degree ~grid. (a) through (g): surface variables; (h) 500 hPa geopotential height; (i) through (l), (m) through (p): temperature, specific humidity, and $u$, $v$ wind at 850 hPa and 250 hPa, respectively. Bars indicate the ensemble medians and error bars indicate the ensemble ranges.}
\label{fig:rmsb_summary}
\end{figure*}

The bias of the AIWCMs relative to ERA5 is a basic metric of climate skill. Fig. \ref{fig:bias_map} shows single ensemble member biases over the training and test period for 2-meter air temperature and surface precipitation, computed on the 1\degree ~resolution grid. (See Appendix \ref{app_2p8deg_results} for the equivalent figure on the 2.8\degree ~grid). The AIWCMs exhibit a range of bias patterns, but typically have larger temperature biases over land and sea ice than over ocean. This is expected given the specified SST AIMIP experiment, and is consistent with the bias patterns of the conventional GFDL-CM4 model (Fig. \ref{fig:bias_map}a, bottom row), which has larger biases over land and sea ice than over ocean. Most of the AIWCMs systematically underpredict temperature in the test period (Fig. \ref{fig:bias_map}b). The precipitation bias patterns (for models that predicted precipitation, Fig. \ref{fig:bias_map}c, d) vary significantly in terms of sign and placement near the equator, where CM4 also has its largest biases.

Figure \ref{fig:rmsb_summary} presents global-averaged RMSB for many variables, computed on the 1 \degree grid, as metrics of climate bias. The AIWCMs are typically able to produce lower RMSB versus ERA5 than the GFDL-CM4 model.  Some AIWCMs consistently produce lower RMSB than others, both at the surface and in the upper atmosphere. The AIWCMs also tend to have higher biases in the test period than in the training period, though some models generalize better than others in this way. In general, the 5-member ensemble variability (black errorbars) for individual AIWCMs is small relative to the magnitude of the biases, though it is larger for cBottle1.3, and for most models higher in the atmosphere.  Figs. \ref{app_fig:bias_map_train_tas_ensemble} and \ref{app_fig:bias_map_test_tas_ensemble} show bias maps for 2-meter air temperature across model ensemble members in the training and test periods. Additional models biases across pressure levels are shown in Appendix \ref{additional_biases}.

The AIWCMs' diversity of model outputs makes some direct comparisons difficult. For example, some models submitted surface pressure, whereas others submitted mean sea level pressure, resulting in different bias magnitudes (Fig. \ref{fig:rmsb_summary}d, e), as surface pressure variability over topography is much larger. Not all models submitted surface precipitation and near-surface humidity as outputs. 

\subsection{E2: Trends}

\begin{figure*}[h]
\includegraphics[width=17.4cm]{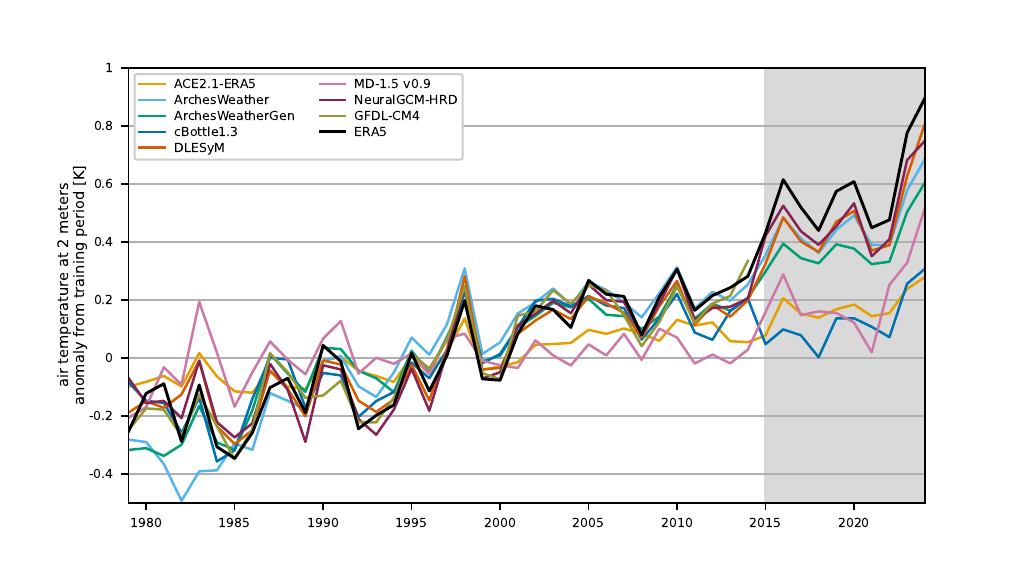}
\caption{Global- and annual-mean 2-meter air temperature, shown as anomalies from the training period (1979-2014) average. ERA5 is in black; AIWCM model ensemble means are shown, along with the CMIP6 GFDL-CM4 single-member prediction. The AIMIP test period (2015-2024) is shaded at right.}
\label{fig:trend_tas}
\end{figure*}

We compute trends first by computing global area-weighted annual mean series, and then fitting linear trends to the training and test periods. Fig. \ref{fig:trend_tas} shows the global- and annual-mean series of 2-meter air temperature for the AIWCM ensemble means, ERA5, and CM4. (The individual ensemble members are plotted in Fig. \ref{app_fig:trend_tas_ens}). Plotting these as an anomaly from the 1979-2014 training period mean for each model emphasizes that some AIWCMs (NeuralGCM, ArchesWeather and ArchesWeatherGen, and DLESyM) more accurately capture ERA5's elevated warming in the 2015-2024 test period than others (ACE2.1-ERA5, MD-1.5 v0.9, cBottle1.3). This is consistent with the difference in mean bias pattern between training and test periods for those models in Fig. \ref{fig:bias_map}a, b. ACE2.1-ERA5 and MD-1.5 v0.9 also struggle more than other models to capture the training period warming trend. GFDL-CM4 tracks the training period trend well, but is unavailable for the test period. 

\begin{figure*}[h]
\includegraphics[width=17.4cm]{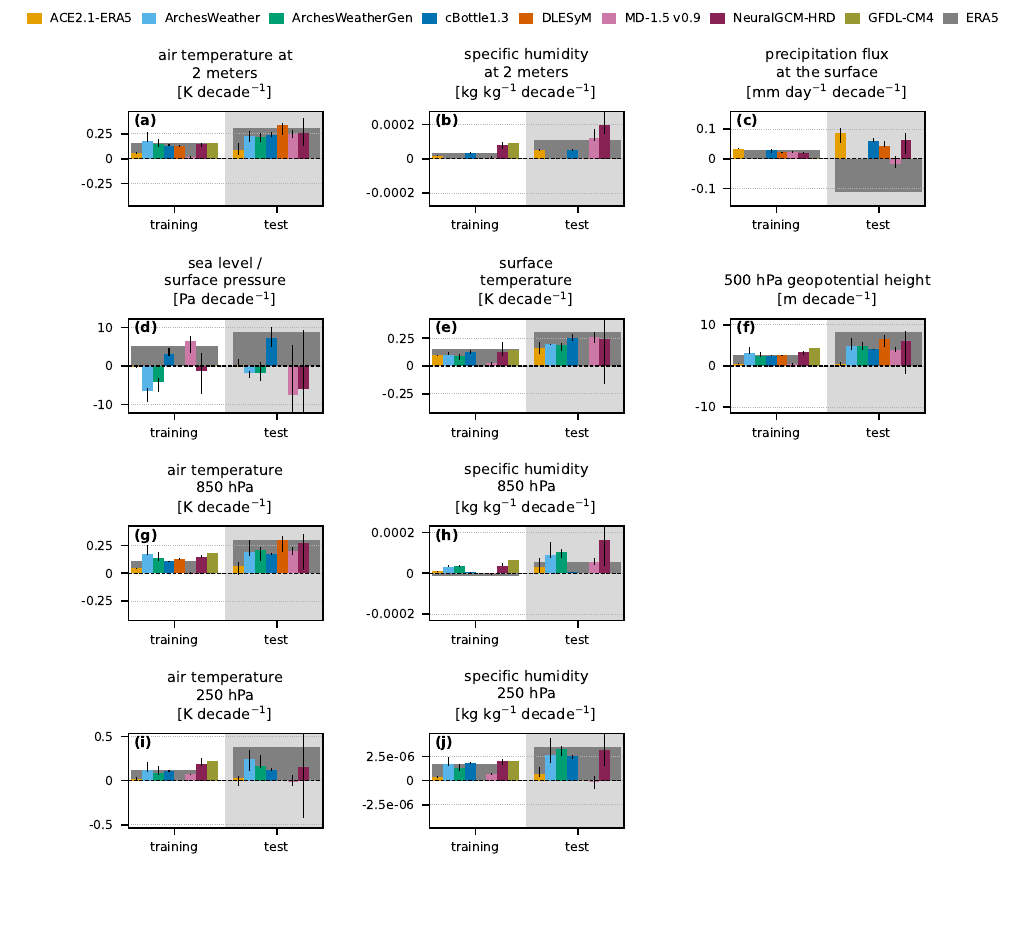}
\caption{Trends of global- and annual-mean variables. (a through e) surface variables, (f) 500 hPa geopotential height, (g), (h) 850 hPa temperature and humidity, and (i), (j) 250 hPa temperature and humidity. In (d) mean sea level pressure trend is shown for all models that submitted this variable, but for ACE2.1-ERA5, MD-1.5 v0.9 and NeuralGCM surface pressure trend is shown. The dark background bar is ERA5. GFDL-CM4 trends are unavailable in the test period. }
\label{fig:trend_bar}
\end{figure*}

\begin{figure*}[h]
\includegraphics[width=17.4cm]{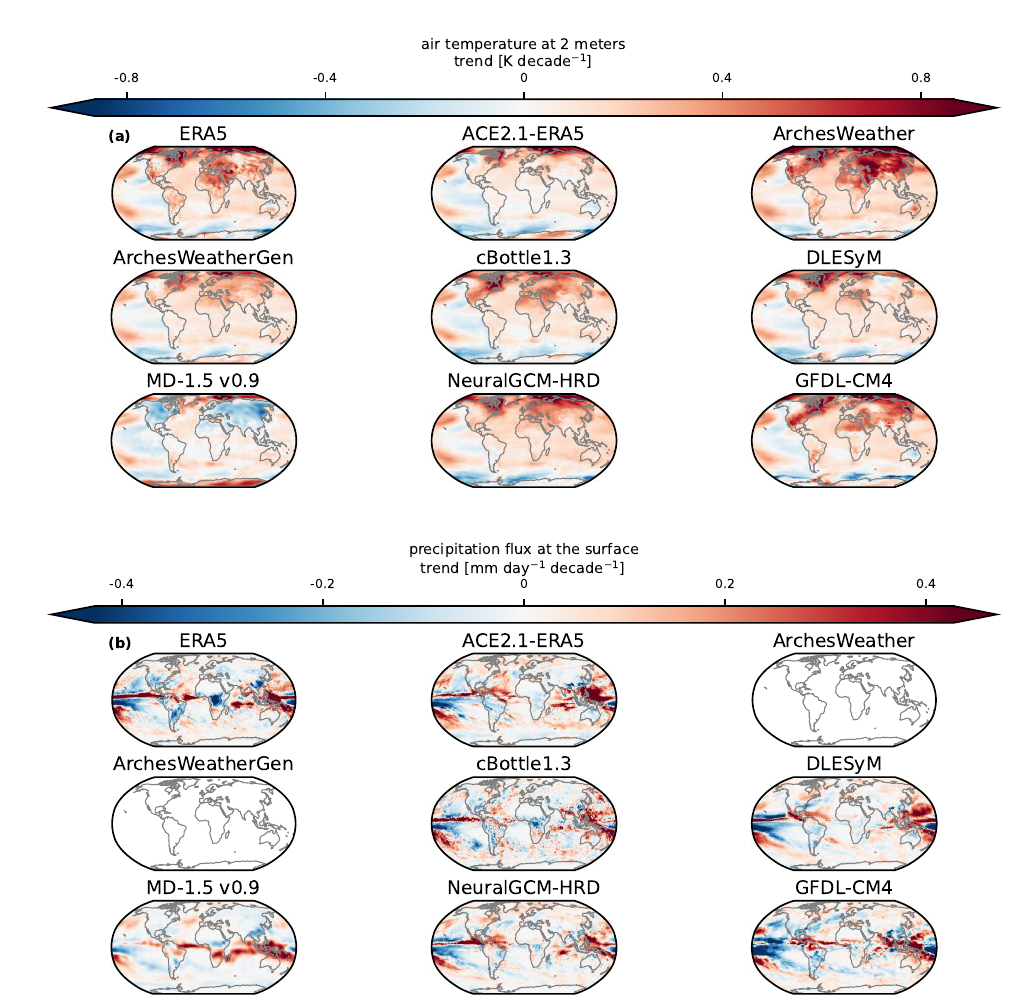}
\caption{Trend maps at 1\degree ~resolution over the training period for (a) 2-meter temperature and (b) surface precipitation.}
\label{fig:trend_maps}
\end{figure*}

Figure \ref{fig:trend_bar} shows global-mean training and test trends for variables that have warming-driven trends in ERA5, though for some variables this is more robustly captured in the 36-year training period than in the 10-year test period. For most variables and models, the AIWCMs correctly reproduce the sign of the training period trends, whereas the test period trend predictions are noisier. However, there is a tendency for most AIWCMs to underestimate the magnitude of the trends to a some degree, even in the training period. Additional global-mean trends across pressure levels are shown in Appendix \ref{additional_trends}. 

We also show maps of trends computed at the gridpoint scale. In Fig. \ref{fig:trend_maps}, we show 2-meter temperature and surface precipitation trends over the training period. The Arctic and land regions have the strongest warming trends in ERA5. The AIWCM trends tend to underestimate the observed Arctic warming maxima slightly and in many cases underestimate the enhanced warming over land, but more notably they produce trend error patterns unique to the individual model (training and test period bias patterns are fairly consistent across Figs. \ref{app_fig:bias_map_train_tas_ensemble} and \ref{app_fig:bias_map_test_tas_ensemble}). The GFDL-CM4 temperature trend pattern error magnitude are comparable to those of the AIWCMs that have smaller trend pattern errors (NeuralGCM, cBottle1.3, DLESyM, ArchesWeatherGen); see also Fig. \ref{fig:trend_bar}a. For precipitation (Fig. \ref{fig:trend_maps}b), the observed ERA5 trend of an intensified intertropical convergence zone in the central Pacific Ocean, and reduced subtropical precipitation, is largely not replicated by the AIWCMs. GFDL-CM4 also does not fully capture this feature, but that may not be a bias of this physically-based model, since ERA5 precipitation trends do not fully track those in observational datasets \citep[e.g., GPCP,][]{GPCP}, which would be a more appropriate reference for CMIP models.

The trends here are computed on the 1\degree ~resolution grid; the equivalent trends computed on the 2.8\degree ~grid, and showing NeuralGCM instead of NeuralGCM-HRD, are shown in Appendix \ref{app_2p8deg_results}. 

\subsection{E3: ENSO response}

\begin{figure*}[h]
\includegraphics[width=17.4cm]{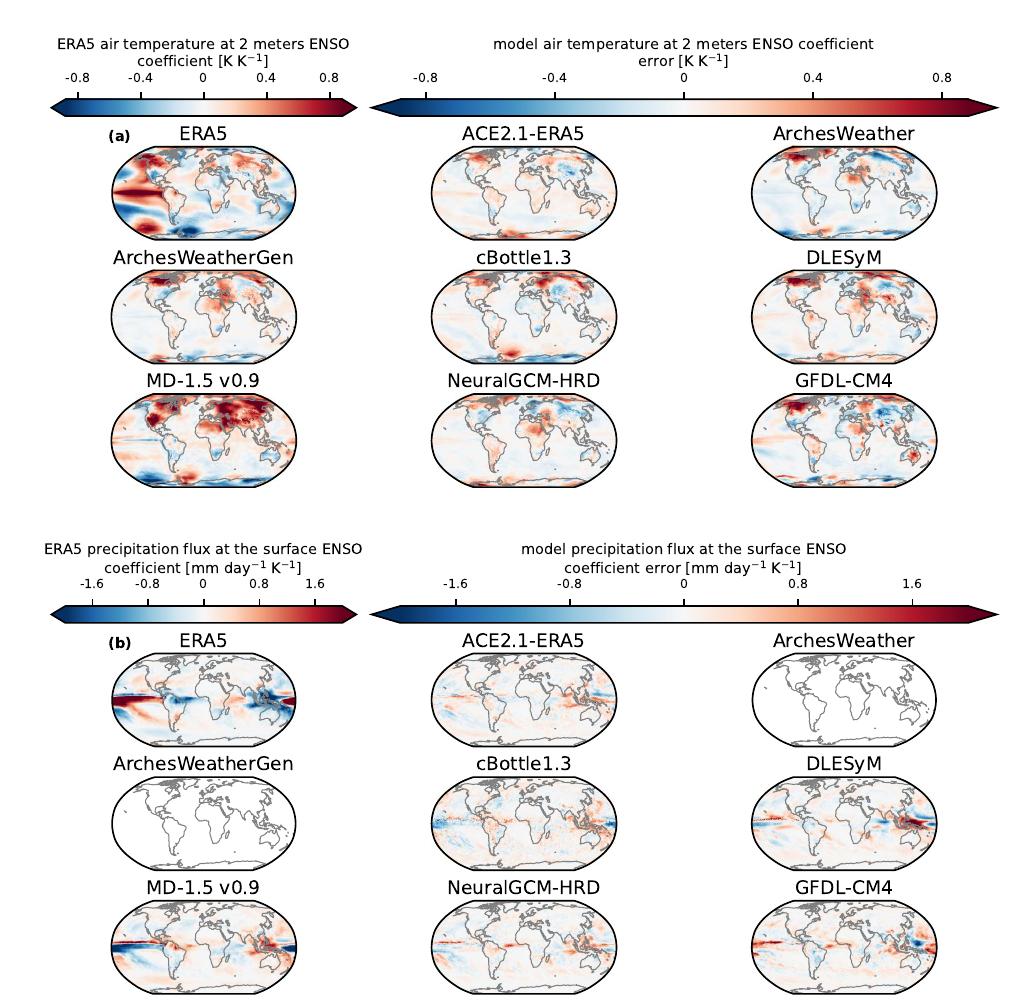}
\caption{ENSO coefficient maps at 1\degree ~resolution for ERA5 (upper left panels) and model coefficient errors versus ERA5 coefficients (subsequent panels) over the training period, for (a) 2-meter temperature and (b) surface precipitation.}
\label{fig:enso_maps}
\end{figure*}

We compute the regression coefficients of variables against the AIMIP-specified Ni\~no3.4 index, using monthly data over the training and test periods. Figure \ref{fig:enso_maps} (first panel) shows 1\degree ~resolution coefficient maps of ERA5 2-meter temperature and surface precipitation over the training period, reflecting the conventional response to Ni\~no3.4 variability, with the greatest coefficient values over the tropical Pacific. The test period maps (not shown) have greater values over extra-tropical land, but this is likely due to insignificant correlations owing to the short test period (2014-2025) relative to the typical ENSO period.

The AIWCMs' ENSO coefficient errors versus ERA5 are shown in subsequent panels in Fig. \ref{fig:enso_maps}, as are those for GFDL-CM4. The AIWCMs' temperature coefficient errors (Fig. \ref{fig:enso_maps}a) are small over tropical oceans relative to the ERA5 coefficient magnitudes, suggesting that the response to the forced variability is broadly learned by the models. For precipitation (Fig. \ref{fig:enso_maps}b), the training period errors are also small (and grow modestly in the test period). 

Global-mean ENSO coefficient errors for a range of variables over the training and test periods are shown in Fig. \ref{app_fig:enso_rmse}. For this broader set of variables, the same patterns hold: most AIWCMs have similar magnitude coefficient errors to one another, though MD-1.5 v0.9 errors tend to be somewhat higher. The test period errors are higher than the training period errors, but by a margin that is fairly consistent across models, likely due to the challenge of computing robust coefficients in the short test period. Relative to CM4, most AIWCMs produce slightly smaller ENSO coefficient errors against ERA5. This is consistent with Fig. \ref{fig:rmsb_summary} in which the AIWCMs typically produce smaller biases than CM4, despite being forced by approximately the same SSTs. 

The equivalent maps of 2.8\degree ~resolution ENSO coefficients are shown in Appendix \ref{app_2p8deg_results}.

\subsection{E4: Daily variability}

We compute the daily variability magnitude of the AIWCMs over the first year of the AIMIP simulations (1979, excluding the 3-month spinup period). This metric is computed as the standard deviation of daily anomalies from the monthly mean values of the same model realization. The AIWCMs vary in their internal prediction timestep (variously using hourly, 3-hourly, and 6-hourly predictions), which may influence their ability to capture the daily average variability evaluated here. MD-1.5 v0.9 makes predictions only at a monthly timestep and is not included.

\begin{figure*}[h]
\includegraphics[width=17.4cm]{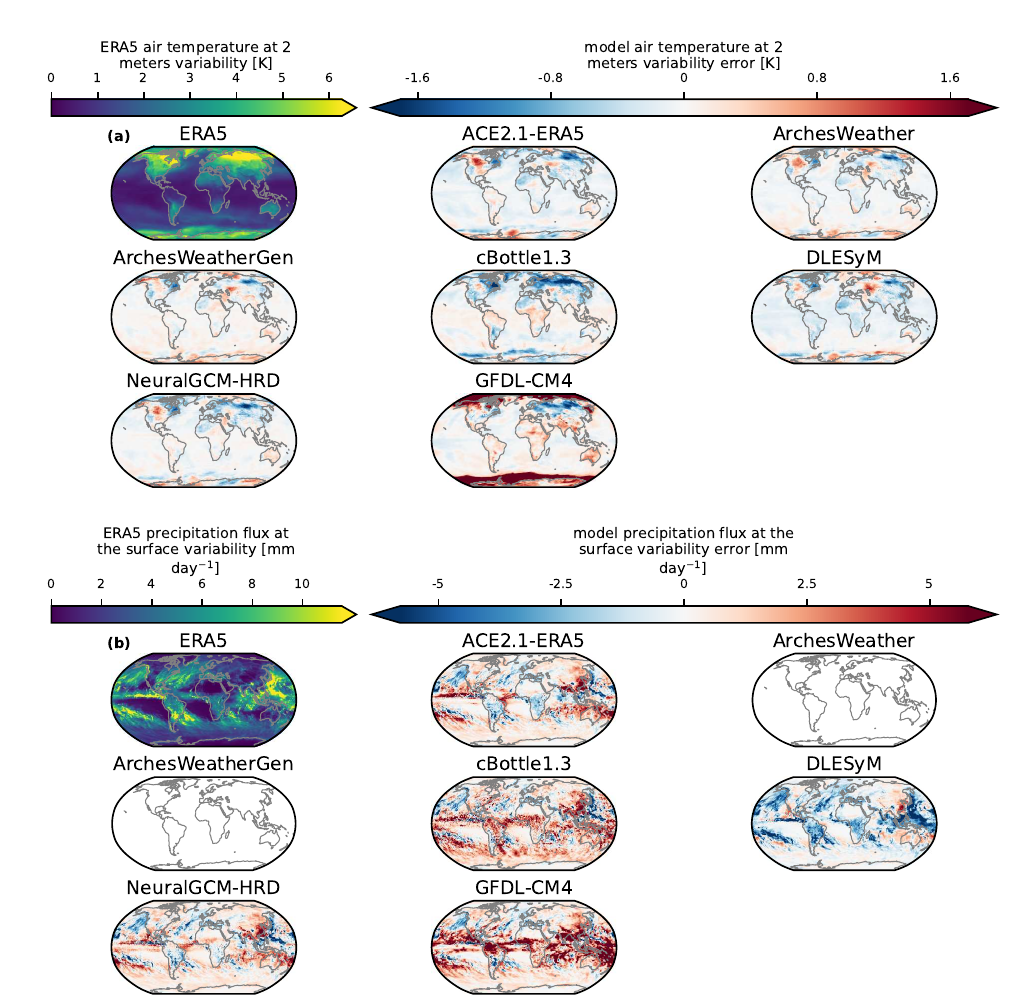}
\caption{Standard deviation of daily anomalies from monthly mean at 1\degree ~resolution over 1979, for 2-meter air temperature (a) and surface precipitation (b). Upper left panels shows anomaly standard deviation in ERA5, and subsequent panels show the error in model anomaly standard deviations.}
\label{fig:daily_variability_error_map}
\end{figure*}

In Fig. \ref{fig:daily_variability_error_map}, we show at 1\degree ~resolution the ERA5 daily anomaly magnitudes of 2-meter air temperature and surface precipitation, along with the model errors of these magnitudes. The 2-meter temperature anomaly magnitude errors are smaller over the ocean than on land because the sea-surface temperature is specified and only slowly varying (Fig. \ref{fig:daily_variability_error_map}a). Some models underestimate the daily anomaly magnitude over ocean, which may be due to the AIMIP forcing being interpolated from monthly mean values (see Appendix \ref{appendix:aimip_forcing}), whereas the evaluation is against ERA5 daily temperature. Nonetheless, some models such as the generative ArchesWeatherGen and cBottle1.3 avoid daily anomaly magnitude biases over tropical ocean. Over land, there are larger errors. The largest daily anomaly values, over northern Eurasia and northeastern North America, tend to be underestimated. The GFDL-CM4 model overestimates daily temperature variability relative to ERA5 over sea ice.

\begin{figure*}[h]
\includegraphics[width=17.4cm]{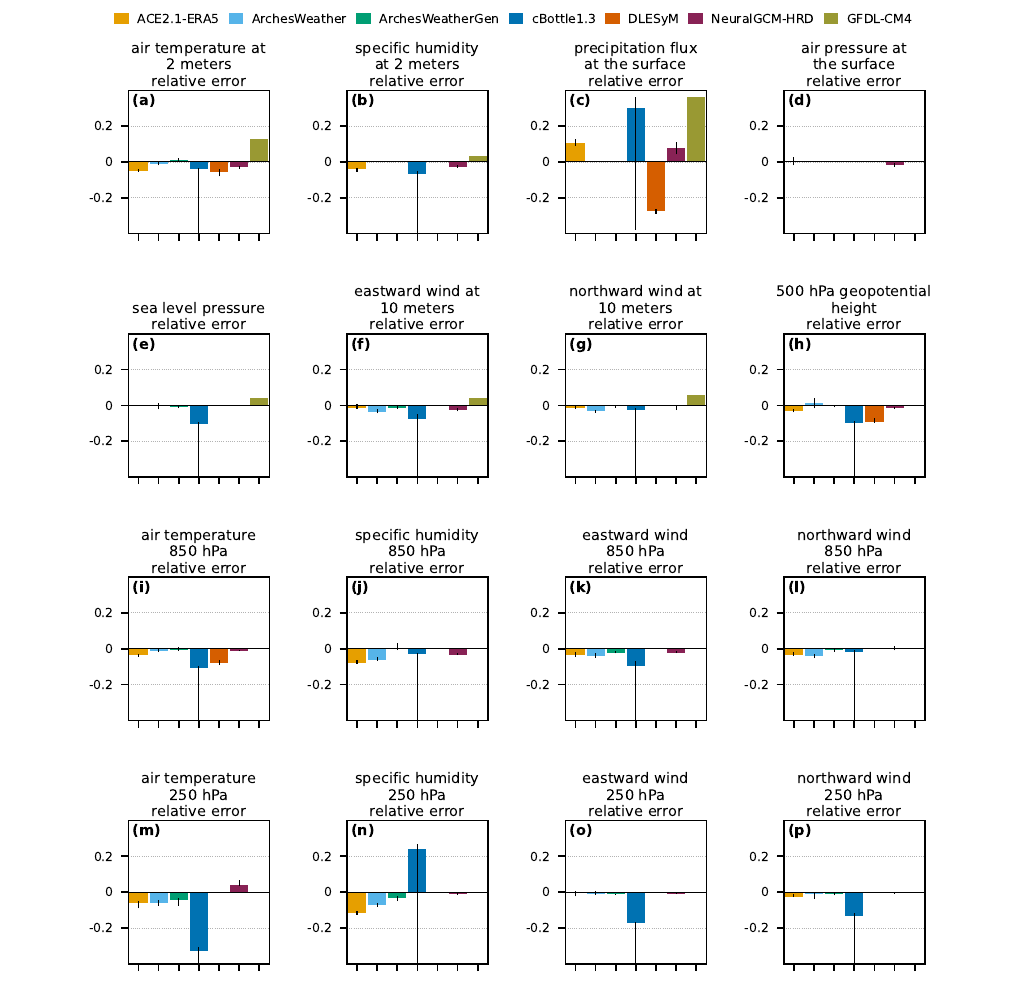}
\caption{Global area-weighted mean of model daily anomaly standard deviation errors, relative to global-mean ERA5 daily variability, at 1\degree ~resolution for the set of variables shown in Fig. \ref{fig:rmsb_summary}. CMIP6 GFDL-CM4 daily variables are not available for pressure-level variables.}
\label{fig:daily_variability_error_bar}
\end{figure*}

Precipitation variability in ERA5, the subset of AIWCMs that provided precipitation as an output (ACE2.1-ERA5, cBottle, DLESyM, and NeuralGCM-HRD), and CM4 are shown in Fig. \ref{fig:daily_variability_error_map}b. The greatest precipitation daily anomaly magnitudes occur in the tropical Pacific Ocean and along mid-latitude oceanic storm tracks. The AIWCM errors in estimating these variability patterns are significant and differ between models, but all models (including CM4) overestimate the daily anomaly magnitude over most wet regions, with the exception of DLESyM. The biases in precipitation variability are possibly related to biases in prediction of the `dry-day fraction' in the AIWCMs relative to ERA5 (see Appendix \ref{additional_daily_variability}), even in this in-sample period.

Over a larger set of variables (Fig. \ref{fig:daily_variability_error_bar}), we see that (except for precipitation) there is a tendency of the AIWCMs to underestimate the daily anomaly magnitude of most variables, by an amount that varies widely but is typically on the order of 2-10\% of the ERA5 daily variability anomaly magnitude. The generative model ArchesWeatherGen tends to have the smallest underestimation of this variability. The conditional diffusion model cBottle1.3 has a wide range of daily variability across its physics ensemble members, which is expected due to the different degree of temporal autocorrelation among ensemble members (member 5 vs. members 1-4, see Sect. \ref{cBottle}). Nonetheless, its median realization (represented by the bars in Fig. \ref{fig:daily_variability_error_bar}) tends toward underestimation of daily variability. In comparison, GFDL-CM4 overestimates the daily variability of many variables relative to ERA5.

\subsection{E5: Perturbed SST response}

\begin{figure*}[h]
\includegraphics[width=17.4cm]{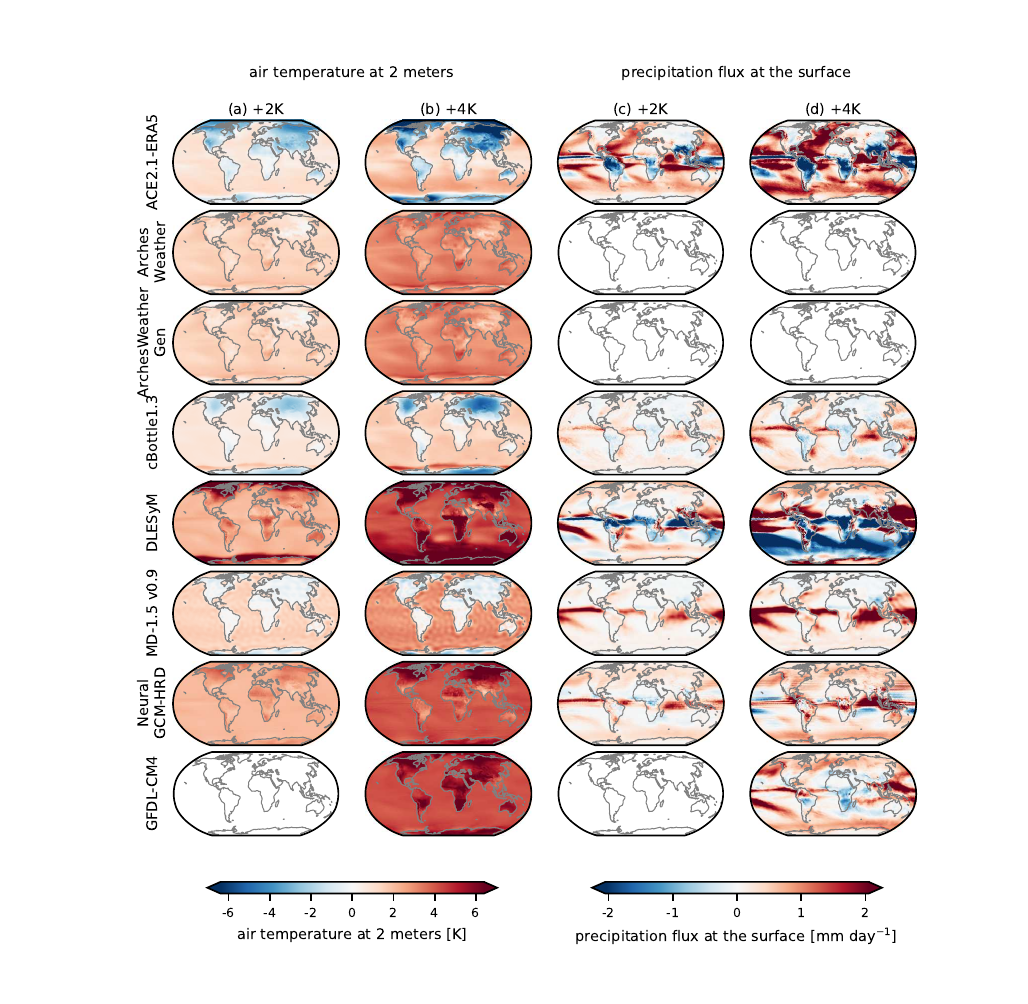}
\caption{Time-mean response to $+2 ~K$ and $+4 ~K$ SST perturbations, for 2-meter air temperature (a), (b) and surface precipitation (c), (d), respectively. Only $+4 ~K$ SST perturbations are available for the GFDL-CM4 model.}
\label{fig:perturbation_response_map}
\end{figure*}

We present both $+2 ~K$ and $+4 ~K$ perturbed SST simulated responses in AIMIP,  computed as the time-mean differences in variables between the perturbed SST and control runs over the simulation (Oct. 1 1978 through Dec. 31 2024). Fig. \ref{fig:perturbation_response_map} shows the 2-meter temperature and surface precipitation responses for each AIWCM, and for GFDL-CM4 for $+4 ~K$ only. The equivalent figure for 2.8\degree ~resolution is in Appendix \ref{app_2p8deg_results}.

The range of responses to these SST perturbations is much larger than for the previous comparisons forced by the SSTs during the training and test periods. This is exactly what one might expect for a strongly out-of-sample test of AI-based models, although it is problematic for modeling of climate change. For 2-meter temperature (Fig. \ref{fig:perturbation_response_map}a, b), some AIWCMs, such as NeuralGCM-HRD and DLESyM, fairly closely match CM4's physically-expected \citep[e.g.,][]{Sutton2007, Byrne2013} results of 2-meter temperature warming that matches the SST perturbation, and enhanced warming over land. Other models have more generalized warming that omit the land enhancement (ArchesWeather and ArchesWeatherGen). However, ACE2.1-ERA5, cBottle1.3, and MD-1.5 v0.9 all both underestimate the ocean warming magnitude and implausibly predict cooling over land. For precipitation response (Fig. \ref{fig:perturbation_response_map}c, d), there is similar variability among AIWCMs, though most seem to agree with CM4's prediction of enhanced precipitation at tropical convergence zones. Developing AI models that can reliably predict future climate trends using historical information and reliable physical knowledge is a key challenge for the AIWCM community over the next few years.  

\section{Discussion}
\label{sec:discussion}

The AIWCMs considered in AIMIP Phase 1 perform quite similarly across a number of metrics, and have some important strengths. Most can, for example, produce lower climate biases versus ERA5 reanalysis (against which they were trained) than a physically-based CMIP6 model (GFDL-CM4), which of course has no direct tuning against ERA5. For other metrics, such as the ENSO response to the prescribed SSTs and the degree of replicating variables' daily variability, the AIWCMs consistently capture most of the target pattern as well. The consistency for these metrics is noteworthy given that the AIWCMs implement a range of frameworks (autoregressive full emulation and hybrid physics/AI models as well as conditional diffusion), and AI architectures (U-Nets, Fourier operators, vision transformers, etc.). 

However, for some metrics, there is significant divergence between the AIWCMs: the ability to capture both training period and test period warming trends varies significantly across models (e.g., Fig. \ref{fig:trend_tas}). Some AIWCMs underpredict the trends in air temperature and other variables that should result from the increasing SSTs. In the particular case of ACE2.1-ERA5, the inability to capture the warming trend is unlike the behavior of a similar version of the model run in an AMIP-type experiment \citep{Watt-Meyer2025}. The difference is likely attributable to the exclusion of CO${_2}$ as forcing in AIMIP, as CO${_2}$ and warming trends are strongly correlated in the historical record.

The ability to produce physically plausible responses to extreme out-of-sample perturbed SST experiments also widely varies between models. It is possible that some of the differences in the AIWCMs' responses may be due to how the prescribed SSTs, which originally contain missing values over land and sea ice, are provided to the AI models as conditioning or forcing, including when spatially-uniform perturbations are applied. The models' handling of this falls into several categories:
\begin{itemize}
    \item Constant filling: A constant value, frequently the climatological mean, is used to fill the land and sea ice SSTs (ArchesWeather/ArchesWeatherGen, cBottle1.3, MD-1.5 v0.9). The land and sea ice SSTs are not changed from this fill value in the perturbed SST experiments.
    \item Interpolation of SSTs: DLESyM zonally interpolates SSTs from the ocean over land and sea ice every 16th timestep. Additionally in this case, the land and sea ice SSTs are perturbed, along with the ocean, in the perturbed SST experiments.
    \item Merging of ocean and land/sea ice fields: SSTs over ocean cells are merged with other values over land and sea ice and no filling or interpolation is required (ACE2.1-ERA5, NeuralGCM/NeuralGCM-HRD).
\end{itemize}

This grouping does not necessarily explain the differences in behavior seen in Fig. \ref{fig:perturbation_response_map}, however, suggesting that other factors such as the AI architecture and its sensitivity to input perturbations at large spatial scales must also play a role. Regardless, a primary goal of climate models is to provide guidance for unseen scenarios, a task for which physically-based models are explicitly built, but one that may challenge AI-based models. These results suggest that AIWCMs ability to achieve this goal is currently a work in progress. 

AIMIP Phase 1 demonstrates that the goal of AI model intercomparison around a common experimental specification and data output format is reasonable to achieve, despite the diverse training methods used to produce the AIWCMs. We hope that this encourages the community to continue these intercomparison efforts and envision future phases of AIMIP, which may include more complex procedures than specified-SST AMIP-type experiments (e.g., coupled earth system component model experiment such as those in CMIP), more extensive data outputs, and more evaluation metrics. The results from this intercomparison may also inform future development of AIWCMs.

\conclusions  

We present the AIMIP Phase 1 project, an AI weather and climate model intercomparison. This paper defines the specifications of AIMIP Phase 1, including the specified SST and SIC experiment modeled after the CMIP's atmospheric model intercomparison project, AMIP. It describes how the AIMIP Phase 1 AIWCMs are to be trained on ERA5 reanalysis data from 1979 to 2014, and are forced only with specified SST, SIC, and solar insolation. The AIMIP Phase 1 simulations are 5-member ensembles that start on Oct. 1, 1978 and run through Dec. 31, 2024, which includes a 10-year test (out-of-sample) period at the end of the simulation. Modeling groups must submit monthly data for a set of surface and pressure-level variables for the entire simulation, as well as daily data for the first 15 months and last 12 months of the simulations. These outputs must be formatted according to CMIP conventions to facilitate common evaluation using existing tools. Additionally, perturbed SST simulation results can be submitted.

We describe the eight AIWCM submissions received by AIMIP Phase 1 from six modeling groups, which span a range of simulation frameworks and AI architectures. Then, we present evaluations of the AIWCMs across five basic metrics that draw on the CMIP conventions: biases, trends, ENSO response to specified SSTs, daily variability magnitudes, and out-of-sample perturbed SST response. We emphasize that the AIMIP Phase 1 model submissions represent snapshots of ongoing model development, and as a result should not be understood as fixed references for the models.

We find that the intercomparison yielded similarities across AIWCMs, including their ability to more faithfully replicate the ERA5 climate patterns than a conventional CMIP6 model, and their strong ability to simulate the ENSO response to specified SSTs. However, we also find that there is a wide range of behavior in the AIWCMs' ability to simulate both the in-sample (training period) and out-of-sample warming trends in ERA5. We note major disagreement in the models' response to the very much out-of-sample perturbed SST experiments. 

We present the AIMIP Phase 1 dataset for further evaluation and experimentation, and hope that by facilitating an intercomparison with a common experiment and data format, the community is able to further leverage the rapid advances in AI model development to achieve more robust and efficient climate modeling tools.




\codedataavailability{

The evaluation code used to produce all figures and analyses in this paper is archived on Zenodo at \url{https://doi.org/10.5281/zenodo.21302180} \citep[][Apache License 2.0]{henn_2026_21302180}. The AIMIP Phase 1 processed data required to reproduce all figures and analyses in this paper are also available in \cite[][CC-BY 4.0 License]{henn_2026_21302180}. For convenience, the full AIMIP Phase 1 output dataset is additionally accessible via an S3 endpoint hosted by DKRZ, with download instructions available in \citep{henn_2026_21302180} and at \url{https://github.com/ai2cm/AIMIP} (last access: 23 July 2026). The AIMIP Phase 1 SST and SIC forcing dataset is archived on Zenodo at \url{https://doi.org/10.5281/zenodo.17065758} \citep[][CC-BY 4.0 License]{arcomano_2025_17065758}.

The training and inference code of each participating model is archived on Zenodo as follows: ACE2.1-ERA5 at \url{https://doi.org/10.5281/zenodo.19831256} \citep[][Apache 2.0 License]{brian_henn_2026_19831256}; ArchesWeather and ArchesWeatherGen at \url{https://doi.org/10.5281/zenodo.20784771} \citep[][BSD 3-Clause License]{guillaume_couairon_2026_20784771}; cBottle-1.3 at \url{https://doi.org/10.5281/zenodo.20832634} \citep[][NVIDIA Software and Model Evaluation License]{manshausen_2026_20832634}; DLESyM at \url{https://doi.org/10.5281/zenodo.21270137} \citep[][MIT License]{nathanielcresswellclay_2026_21270137}; MD-1.5 v0.9 at \url{https://doi.org/10.5281/zenodo.21430292} \citep[][MIT License]{kyle_hall_2026_21430292}; and NeuralGCM and NeuralGCM-HRD at \url{https://doi.org/10.5281/zenodo.21303721} \citep[][Apache 2.0 License]{kochkov_2026_21303721}. See Table \ref{tab:submissions} for more details on the participating models.

The CMIP6 GFDL-CM4 output used for comparison — the amip experiment (activity CMIP: monthly Amon variables hus, huss, pr, ps, psl, ta, tas, ts, ua, uas, va, vas, and zg; daily day variables huss, pr, psl, tas, uas, and vas) and the amip-p4K experiment (activity CFMIP: monthly Amon variables huss, pr, ta, tas, ts, and zg), each for variant r1i1p1f1, grid label gr1, dataset version v20180701 — is archived and distributed by the Earth System Grid Federation at \url{https://doi.org/10.22033/ESGF/CMIP6.8494} \citep{https://doi.org/10.22033/ESGF/CMIP6.8494} and \url{https://doi.org/10.22033/ESGF/CMIP6.8508} \citep{https://doi.org/10.22033/ESGF/CMIP6.8508}; for this study these data were accessed as analysis-ready, cloud-optimized Zarr stores via the ESGF/Pangeo CMIP6 Google Cloud collection (\url{https://pangeo-data.github.io/pangeo-cmip6-cloud}, last access: 23 July 2026).

The ERA5 reanalysis (Hersbach et al., 2020) used for model training and as the evaluation reference is available from the Copernicus Climate Change Service Climate Data Store at \url{https://doi.org/10.24381/cds.adbb2d47} and \url{https://doi.org/10.24381/cds.bd0915c6} \citep{era5_hourly_single_levels, era5_hourly_pressure_levels}.

}



\authorcontribution{CSB, NK, CL, OWM, and MJM helped with the intercomparison specification and data planning. BH conducted the evaluations across models. BH trained and ran simulations with the ACE2.1-ERA5 model; CSB, TA, OWM provided guidance on its training and evaluation. RS, RB, GC, YH, AJ, CL contributed to the ArchesWeather/ArchesWeatherGen submission. RS, RB, GC, YH contributed to the code for adapting ArchesWeather to the AIMIP protocol. RS and RB conducted the rollouts and evaluation of the models. RS ran the ablations for validating the model. RB prepared training and validation data. AJ provided the CMOR code for the project and helped with evaluation. CL provided guidance on the project. PM trained and ran inferences of cBottle1.3, adding the correlated latents feature, PM and NB post processed the data, NB provided guidance on training and inference as well as handling the Open Source release. NCC ran simulations with DLESyM; DD provided guidance on its training and evaluation. KJCH and MJM co-designed the MD-1.5 v0.9 architecture, and KJCH developed the implementation. KJCH trained the submitted model and ran inference for the AIMIP evaluation period. MJM and KJCH wrote the description included here. JY, DK, ILG, SH contributed to NeuralGCM/NeuralGCM-HRD submission. JY, DK and SH contributed to the code for adapting NeuralGCM to the AIMIP protocol (adding surface fields, downscaling to 1 degree etc.). JY trained the model and ran model inference. ILG provided the CMOR code for the project.}

\competinginterests{The authors declare that they have no conflict of interest.} 


\begin{acknowledgements}
Community members submitted valuable input towards the ideation and specification of an AI weather and climate model intercomparison, including Paul Ullrich, Veronika Eyring, and many others. We acknowledge Bettina Gier of University of Bremen for help in regridding and coarsening the ERA5 data for evaluation. Ai2 is supported by the estate of Paul G. Allen. For the CMIP6 data, we acknowledge the World Climate Research Programme, which, through its CMIP activity, coordinated and promoted this model intercomparison project. We thank the climate modeling groups for producing and making available their model output. We specifically acknowledge the Pangeo initiative for providing the Google Cloud storage and infrastructure that facilitated access to this data. NCC was supported by the Collaborative Research in Environmental Science and Sustainability Training (CRESST) program at the University of Washington's College of the Environment. DD was supported by Office of Naval Research grant N00014‐24‐12528. NK was supported by project S1: Diagnosis and Metrics in Climate Models of the Collaborative Research Centre TRR 181 “Energy Transfer in Atmosphere and Ocean”, funded by the Deutsche Forschungsgemeinschaft (DFG, German Research Foundation, project no. 274762653). Computing resources for the DLESyM effort were provided by a Grant from the NVIDIA Applied Research Accelerator Program. KJCH and MJM were supported by the U.S. DOE, Office of Science, Office of Biological and Environmental Research (BER), RGMA component of the Earth and Environmental System Modeling Program under Award \#DE-SC0024093. We gratefully acknowledge DKRZ (the Deutsches Klimarechenzentrum in Hamburg, Germany) for providing S3 storage for AIMIP project.
\end{acknowledgements}








\bibliographystyle{copernicus}
\bibliography{main.bib}

@Article{DunneCMIP7,
AUTHOR = {Dunne, J. P. and Hewitt, H. T. and Arblaster, J. M. and Bonou, F. and Boucher, O. and Cavazos, T. and Dingley, B. and Durack, P. J. and Hassler, B. and Juckes, M. and Miyakawa, T. and Mizielinski, M. and Naik, V. and Nicholls, Z. and O'Rourke, E. and Pincus, R. and Sanderson, B. M. and Simpson, I. R. and Taylor, K. E.},
TITLE = {An evolving Coupled Model Intercomparison Project phase 7 (CMIP7) and Fast Track in support of future climate assessment},
JOURNAL = {Geoscientific Model Development},
VOLUME = {18},
YEAR = {2025},
NUMBER = {19},
PAGES = {6671--6700},
URL = {https://gmd.copernicus.org/articles/18/6671/2025/},
DOI = {10.5194/gmd-18-6671-2025}
}

@article{deser2012uncertainty,
  title={Uncertainty in climate change projections: the role of internal variability},
  author={Deser, Clara and Phillips, Adam and Bourdette, Vincent and Teng, Haiyan},
  journal={Climate dynamics},
  volume={38},
  number={3},
  pages={527--546},
  year={2012},
  publisher={Springer}
}

@misc{mctaggartcowan2026wpmipartificialintelligencehybrid,
      title={WP-MIP: An Artificial Intelligence, Hybrid, and Physically Based Model Intercomparison Project for Weather Prediction}, 
      author={Ron McTaggart-Cowan and Linus Magnusson and Inna Polichtchouk and Duncan Ackerley and Martin Koehler and Barbara Casati and Jan-Huey Chen and Debra Hudson and Masashi Ujiie and Nurizana Amir Aziz and Massimo Bonavita and Zied Ben Bouallegue and Catherine de Burgh-Day and Stephane Chamberland and Kyounngmi Cho and Caio A. S. Coelho and Rostislav Fadeev and Manuel Fuentes and Jorge L. Garcia Franco and Claude Gilbert and Bruno S. Guimaraes and Chris Harris and Michelle Harrold and Syed Husain and Molly James and Alex Kaltenbaugh and Marta Koch and Paulo Y. Kubota and Eun-Hee Lee and Chen Li and Wei Li and Weiwei Li and Llorenc Lledo and Nicholas Loveday and Chrstian Lussana and Zubiar Maalick and Mohau J. Mateyisi and Amy McGovern and Koos van der Merwe and Joel Miller and Marion Mittermaier and Richard Mladek and Kathryn Newman and Andre L. O. Neves and John Pill and Roland Potthast and Maheswar Pradhan and Subhrajit Rath and David S. Richardson and Leo Separovic and Michelle Simoes Reboita and Gregor Skok and Ankur Srivastava and Mikhail Tolstykh and Zhuo Wang and Beth J. Woodham and Fanglin Yang and Radomir Zaripov and Gan Zhang and Hongyan Zhu},
      year={2026},
      eprint={2604.16643},
      archivePrefix={arXiv},
      primaryClass={physics.ao-ph},
      url={https://arxiv.org/abs/2604.16643}, 
}

@misc{AMIP_forcing,
  author       = {Taylor, Karl E. and Williamson, David and Zwiers, Francis},
  title        = {AMIP Sea Surface Temperature and Sea Ice Concentration Boundary Conditions},
  year         = {1997},
  url          = {https://pcmdi.llnl.gov/mips/amip/details/index.html},
  note         = {Accessed: 2024-04-01}
}

@article{Eyring2016,
   author = {Veronika Eyring and Sandrine Bony and Gerald A. Meehl and Catherine A. Senior and Bjorn Stevens and Ronald J. Stouffer and Karl E. Taylor},
   doi = {10.5194/gmd-9-1937-2016},
   issn = {1991-9603},
   issue = {5},
   journal = {Geoscientific Model Development},
   month = {5},
   pages = {1937-1958},
   publisher = {Copernicus GmbH},
   title = {Overview of the Coupled Model Intercomparison Project Phase 6 (CMIP6) experimental design and organization},
   volume = {9},
   url = {https://gmd.copernicus.org/articles/9/1937/2016/},
   year = {2016}
}

@article{
Couairon2026,
author = {Guillaume Couairon  and Renu Singh  and Anastase Charantonis  and Christian Lessig  and Claire Monteleoni },
title = {ArchesWeatherGen: Skillful and compute-efficient probabilistic weather forecasting with machine learning},
journal = {Science Advances},
volume = {12},
number = {17},
pages = {eadx2372},
year = {2026},
doi = {10.1126/sciadv.adx2372},
URL = {https://www.science.org/doi/abs/10.1126/sciadv.adx2372},
eprint = {https://www.science.org/doi/pdf/10.1126/sciadv.adx2372}}

@article{Watt-Meyer2025,
   author = {Oliver Watt-Meyer and Brian Henn and Jeremy McGibbon and Spencer K. Clark and Anna Kwa and W. Andre Perkins and Elynn Wu and Lucas Harris and Christopher S. Bretherton},
   doi = {10.1038/s41612-025-01090-0},
   issn = {2397-3722},
   issue = {1},
   journal = {npj Climate and Atmospheric Science},
   month = {5},
   pages = {205},
   title = {ACE2: accurately learning subseasonal to decadal atmospheric variability and forced responses},
   volume = {8},
   url = {https://www.nature.com/articles/s41612-025-01090-0},
   year = {2025}
}

@article{Kochkov2024,
   author = {Dmitrii Kochkov and Janni Yuval and Ian Langmore and Peter Norgaard and Jamie Smith and Griffin Mooers and Milan Klöwer and James Lottes and Stephan Rasp and Peter Düben and Sam Hatfield and Peter Battaglia and Alvaro Sanchez-Gonzalez and Matthew Willson and Michael P. Brenner and Stephan Hoyer},
   doi = {10.1038/s41586-024-07744-y},
   issn = {0028-0836},
   issue = {8027},
   journal = {Nature},
   month = {8},
   pages = {1060-1066},
   title = {Neural general circulation models for weather and climate},
   volume = {632},
   url = {https://www.nature.com/articles/s41586-024-07744-y},
   year = {2024}
}

@article{Yuval2026,
   author = {Janni Yuval and Ian Langmore and Dmitrii Kochkov and Stephan Hoyer},
   doi = {10.1126/sciadv.adv6891},
   issn = {2375-2548},
   issue = {2},
   journal = {Science Advances},
   month = {1},
   pages = {1060-1066},
   title = {Neural general circulation models for modeling precipitation},
   volume = {12},
   url = {https://www.science.org/doi/10.1126/sciadv.adv6891},
   year = {2026}
}

@misc{brenowitz2025climatebottlegenerativefoundation,
      title={Climate in a Bottle: Towards a Generative Foundation Model for the Kilometer-Scale Global Atmosphere}, 
      author={Noah D. Brenowitz and Tao Ge and Akshay Subramaniam and Peter Manshausen and Aayush Gupta and David M. Hall and Morteza Mardani and Arash Vahdat and Karthik Kashinath and Michael S. Pritchard},
      year={2025},
      eprint={2505.06474},
      archivePrefix={arXiv},
      primaryClass={physics.ao-ph},
      url={https://arxiv.org/abs/2505.06474}, 
}

@article{Cresswell-Clay2025,
   author = {Nathaniel Cresswell-Clay and Bowen Liu and Dale R. Durran and Zihui Liu and Zachary I. Espinosa and Raul A. Moreno and Matthias Karlbauer},
   doi = {10.1029/2025AV001706},
   issn = {2576604X},
   issue = {4},
   journal = {AGU Advances},
   month = {8},
   publisher = {John Wiley and Sons Inc},
   title = {A Deep Learning Earth System Model for Efficient Simulation of the Observed Climate},
   volume = {6},
   year = {2025}
}

@article{Hersbach2020,
   author = {Hans Hersbach and Bill Bell and Paul Berrisford and Shoji Hirahara and András Horányi and Joaquín Muñoz-Sabater and Julien Nicolas and Carole Peubey and Raluca Radu and Dinand Schepers and Adrian Simmons and Cornel Soci and Saleh Abdalla and Xavier Abellan and Gianpaolo Balsamo and Peter Bechtold and Gionata Biavati and Jean Bidlot and Massimo Bonavita and Giovanna De Chiara and Per Dahlgren and Dick Dee and Michail Diamantakis and Rossana Dragani and Johannes Flemming and Richard Forbes and Manuel Fuentes and Alan Geer and Leo Haimberger and Sean Healy and Robin J. Hogan and Elías Hólm and Marta Janisková and Sarah Keeley and Patrick Laloyaux and Philippe Lopez and Cristina Lupu and Gabor Radnoti and Patricia de Rosnay and Iryna Rozum and Freja Vamborg and Sebastien Villaume and Jean Noël Thépaut},
   doi = {10.1002/qj.3803},
   issn = {1477870X},
   issue = {730},
   journal = {Quarterly Journal of the Royal Meteorological Society},
   month = {7},
   pages = {1999-2049},
   publisher = {John Wiley and Sons Ltd},
   title = {The ERA5 global reanalysis},
   volume = {146},
   year = {2020}
}

@article{Allan2022GlobalCIA,
  title={Global Changes in Water Vapor 1979–2020},
  author={R. Allan and K. Willett and V. John and T. Trent},
  journal={Journal of Geophysical Research: Atmospheres},
  year={2022},
  volume={127},
  url={https://doi.org/10.1029/2022JD036728}
}

@article{Loeb2022EvaluatingTTA,
  title={Evaluating Twenty‐Year Trends in Earth's Energy Flows From Observations and Reanalyses},
  author={N. Loeb and M. Mayer and S. Kato and J. Fasullo and H. Zuo and Retish Senan and J. Lyman and G. Johnson and M. Balmaseda},
  journal={Journal of Geophysical Research: Atmospheres},
  year={2022},
  volume={127},
  url={https://api.semanticscholar.org/CorpusId:249296351}
}

@article{Gorski_2005,
   title={HEALPix: A Framework for High‐Resolution Discretization and Fast Analysis of Data Distributed on the Sphere},
   volume={622},
   ISSN={1538-4357},
   url={http://dx.doi.org/10.1086/427976},
   DOI={10.1086/427976},
   number={2},
   journal={The Astrophysical Journal},
   publisher={American Astronomical Society},
   author={Gorski, K. M. and Hivon, E. and Banday, A. J. and Wandelt, B. D. and Hansen, F. K. and Reinecke, M. and Bartelmann, M.},
   year={2005},
   month=apr, pages={759–771},
}

@article{Lavers2022,
   author = {David A. Lavers and Adrian Simmons and Freja Vamborg and Mark J. Rodwell},
   doi = {10.1002/qj.4351},
   issn = {1477870X},
   issue = {748},
   journal = {Quarterly Journal of the Royal Meteorological Society},
   month = {10},
   pages = {3152-3165},
   publisher = {John Wiley and Sons Ltd},
   title = {An evaluation of ERA5 precipitation for climate monitoring},
   volume = {148},
   year = {2022}
}

@article{Ullrich2025,
   author = {P. A. Ullrich and E. A. Barnes and W. D. Collins and K. Dagon and S. Duan and J. Elms and J. Lee and L. R. Leung and D. Lu and M. J. Molina and T. A. O’Brien and F. O. Rebassoo},
   doi = {10.1029/2024jh000496},
   issn = {2993-5210},
   issue = {1},
   journal = {Journal of Geophysical Research: Machine Learning and Computation},
   month = {3},
   publisher = {American Geophysical Union (AGU)},
   title = {Recommendations for Comprehensive and Independent Evaluation of Machine Learning‐Based Earth System Models},
   volume = {2},
   year = {2025}
}

@article{GPCP,
    author = {Adler, R.F. and G.J. Huffman and A. Chang and R. Ferraro and P. Xie and J. Janowiak and B. Rudolf and U. Schneider and S. Curtis and D. Bolvin and A. Gruber and J. Susskind and P. Arkin},
    title = {The Version 2 Global Precipitation Climatology Project (GPCP) Monthly Precipitation Analysis (1979-Present)},
    year = {2003},
    journal = {J. Hydrometeor.},
    volume = {4},
    pages = {1147-1167},
}

@article{Webb2017,
   author = {Mark J. Webb and Timothy Andrews and Alejandro Bodas-Salcedo and Sandrine Bony and Christopher S. Bretherton and Robin Chadwick and Hélène Chepfer and Hervé Douville and Peter Good and Jennifer E. Kay and Stephen A. Klein and Roger Marchand and Brian Medeiros and A. Pier Siebesma and Christopher B. Skinner and Bjorn Stevens and George Tselioudis and Yoko Tsushima and Masahiro Watanabe},
   doi = {10.5194/gmd-10-359-2017},
   issn = {1991-9603},
   issue = {1},
   journal = {Geoscientific Model Development},
   month = {1},
   pages = {359-384},
   title = {The Cloud Feedback Model Intercomparison Project (CFMIP) contribution to CMIP6},
   volume = {10},
   year = {2017}
}

@article{Sutton2007,
   author = {Rowan T. Sutton and Buwen Dong and Jonathan M. Gregory},
   doi = {10.1029/2006GL028164},
   issn = {00948276},
   issue = {2},
   journal = {Geophysical Research Letters},
   month = {1},
   title = {Land/sea warming ratio in response to climate change: IPCC AR4 model results and comparison with observations},
   volume = {34},
   year = {2007}
}

@misc{arcomano_2025_17065758,
  author       = {Arcomano, Troy and
                  Henn, Brian and
                  Bretherton, Christopher},
  title        = {AIMIP Phase 1 Forcing Dataset},
  month        = sep,
  year         = 2025,
  publisher    = {Zenodo},
  doi          = {10.5281/zenodo.17065758},
  url          = {https://doi.org/10.5281/zenodo.17065758},
}

@InProceedings{pmlr-v202-bonev23a,
  title = 	 {Spherical {F}ourier Neural Operators: Learning Stable Dynamics on the Sphere},
  author =       {Bonev, Boris and Kurth, Thorsten and Hundt, Christian and Pathak, Jaideep and Baust, Maximilian and Kashinath, Karthik and Anandkumar, Anima},
  booktitle = 	 {Proceedings of the 40th International Conference on Machine Learning},
  pages = 	 {2806--2823},
  year = 	 {2023},
  editor = 	 {Krause, Andreas and Brunskill, Emma and Cho, Kyunghyun and Engelhardt, Barbara and Sabato, Sivan and Scarlett, Jonathan},
  volume = 	 {202},
  series = 	 {Proceedings of Machine Learning Research},
  month = 	 {23--29 Jul},
  publisher =    {PMLR},
  url = 	 {https://proceedings.mlr.press/v202/bonev23a.html}
}

@misc{rombach2022highresolutionimagesynthesislatent,
      title={High-Resolution Image Synthesis with Latent Diffusion Models}, 
      author={Robin Rombach and Andreas Blattmann and Dominik Lorenz and Patrick Esser and Björn Ommer},
      year={2022},
      eprint={2112.10752},
      archivePrefix={arXiv},
      primaryClass={cs.CV},
      url={https://arxiv.org/abs/2112.10752}, 
}

@misc{zhuang_xesmf_2020,
  author       = {Zhuang, Jiawei and others},
  title        = {pangeo-data/xESMF: Universal Regridder for Geospatial Data},
  year         = 2020,
  publisher    = {Zenodo},
  doi          = {10.5281/zenodo.4294774},
  url          = {https://doi.org/10.5281/zenodo.4294774}
}

@misc{hall2026monthlydiffusionv09latent,
      title={Monthly Diffusion v0.9: A Latent Diffusion Model for the First AI-MIP}, 
      author={Kyle J. C. Hall and Maria J. Molina},
      year={2026},
      eprint={2604.13481},
      archivePrefix={arXiv},
      primaryClass={cs.LG},
      url={https://arxiv.org/abs/2604.13481}, 
}

@article{https://doi.org/10.1002/2017MS001208,
author = {Zhao, M. and Golaz, J.-C. and Held, I. M. and Guo, H. and Balaji, V. and Benson, R. and Chen, J.-H. and Chen, X. and Donner, L. J. and Dunne, J. P. and Dunne, K. and Durachta, J. and Fan, S.-M. and Freidenreich, S. M. and Garner, S. T. and Ginoux, P. and Harris, L. M. and Horowitz, L. W. and Krasting, J. P. and Langenhorst, A. R. and Liang, Z. and Lin, P. and Lin, S.-J. and Malyshev, S. L. and Mason, E. and Milly, P. C. D. and Ming, Y. and Naik, V. and Paulot, F. and Paynter, D. and Phillipps, P. and Radhakrishnan, A. and Ramaswamy, V. and Robinson, T. and Schwarzkopf, D. and Seman, C. J. and Shevliakova, E. and Shen, Z. and Shin, H. and Silvers, L. G. and Wilson, J. R. and Winton, M. and Wittenberg, A. T. and Wyman, B. and Xiang, B.},
title = {The GFDL Global Atmosphere and Land Model AM4.0/LM4.0: 1. Simulation Characteristics With Prescribed SSTs},
journal = {Journal of Advances in Modeling Earth Systems},
volume = {10},
number = {3},
pages = {691-734},
doi = {https://doi.org/10.1002/2017MS001208},
url = {https://agupubs.onlinelibrary.wiley.com/doi/abs/10.1002/2017MS001208},
eprint = {https://agupubs.onlinelibrary.wiley.com/doi/pdf/10.1002/2017MS001208},
year = {2018}
}

@misc{https://doi.org/10.22033/ESGF/CMIP6.8494,
      url = {https://doi.org/10.22033/ESGF/CMIP6.8494},
      title = {NOAA-GFDL GFDL-CM4 model output amip},
      publisher = {Earth System Grid Federation},
      year = {2018},
      author = {Guo, Huan and John, Jasmin G and Blanton, Chris and McHugh, Colleen and Nikonov, Serguei and Radhakrishnan, Aparna and Rand, Kristopher and Zadeh, Niki T. and Balaji, V and Durachta, Jeff and Dupuis, Christopher and Menzel, Raymond and Robinson, Thomas and Underwood, Seth and Vahlenkamp, Hans and Bushuk, Mitchell and Dunne, Krista A. and Dussin, Raphael and Gauthier, Paul PG and Ginoux, Paul and Griffies, Stephen M. and Hallberg, Robert and Harrison, Matthew and Hurlin, William and Lin, Pu and Malyshev, Sergey and Naik, Vaishali and Paulot, Fabien and Paynter, David J and Ploshay, Jeffrey and Reichl, Brandon G and Schwarzkopf, Daniel M and Seman, Charles J and Shao, Andrew and Silvers, Levi and Wyman, Bruce and Yan, Xiaoqin and Zeng, Yujin and Adcroft, Alistair and Dunne, John P. and Held, Isaac M and Krasting, John P. and Horowitz, Larry W. and Milly, P.C.D and Shevliakova, Elena and Winton, Michael and Zhao, Ming and Zhang, Rong},
      doi = {10.22033/ESGF/CMIP6.8494}
}

@article{Gates1992,
   author = {W. Lawrence Gates},
   doi = {10.1175/1520-0477(1992)073<1962:ATAMIP>2.0.CO;2},
   issn = {0003-0007},
   issue = {12},
   journal = {Bulletin of the American Meteorological Society},
   month = {12},
   pages = {1962-1970},
   title = {AMIP: The Atmospheric Model Intercomparison Project},
   volume = {73},
   year = {1992}
}

@article{Byrne2013,
   author = {Michael P. Byrne and Paul A. O’Gorman},
   doi = {10.1175/JCLI-D-12-00262.1},
   issn = {0894-8755},
   issue = {12},
   journal = {Journal of Climate},
   month = {6},
   pages = {4000-4016},
   title = {Land–Ocean Warming Contrast over a Wide Range of Climates: Convective Quasi-Equilibrium Theory and Idealized Simulations},
   volume = {26},
   year = {2013}
}

@article{Barnston1997,
   author = {Anthony G. Barnston and Muthuvel Chelliah and Stanley B. Goldenberg},
   doi = {10.1080/07055900.1997.9649597},
   issn = {0705-5900},
   issue = {3},
   journal = {Atmosphere-Ocean},
   month = {1},
   pages = {367-383},
   title = {Documentation of a highly ENSO‐related sst region in the equatorial pacific: Research note},
   volume = {35},
   year = {1997}
}

@article {Gates1999,
      author = "W. Lawrence Gates and James S. Boyle and Curt Covey and Clyde G. Dease and Charles M. Doutriaux and Robert S. Drach and Michael Fiorino and Peter J. Gleckler and Justin J. Hnilo and Susan M. Marlais and Thomas J. Phillips and Gerald L. Potter and Benjamin D. Santer and Kenneth R. Sperber and Karl E. Taylor and Dean N. Williams",
      title = "An Overview of the Results of the Atmospheric Model Intercomparison Project (AMIP I)",
      journal = "Bulletin of the American Meteorological Society",
      year = "1999",
      publisher = "American Meteorological Society",
      address = "Boston MA, USA",
      volume = "80",
      number = "1",
      doi = "10.1175/1520-0477(1999)080<0029:AOOTRO>2.0.CO;2",
      pages=      "29 - 56",
      url = "https://journals.ametsoc.org/view/journals/bams/80/1/1520-0477_1999_080_0029_aootro_2_0_co_2.xml"
}

@misc{mauzey_2024_10946710,
  author       = {Mauzey, Chris and
                  Durack, Paul and
                  Taylor, Karl E. and
                  Florek, Piotr and
                  Doutriaux, Charles and
                  Nadeau, Denis and
                  Hogan, Emma and
                  Kettleborough, Jamie and
                  Weigel, Tobias and
                  kjoti and
                  jmrgonza and
                  Nicholls, Zeb and
                  Betts, Edward and
                  Seddon, Jon and
                  Wachsmann, Fabian},
  title        = {PCMDI/CMOR: CMOR v3.8.0},
  month        = apr,
  year         = 2024,
  publisher    = {Zenodo},
  version      = {v3.8.0},
  doi          = {10.5281/zenodo.10946710},
  url          = {https://doi.org/10.5281/zenodo.10946710},
}

@misc{brian_henn_2026_19831256,
  author       = {Brian Henn and
                  Watt-Meyer, Oliver and
                  Arcomano, Troy and
                  McGibbon, Jeremy and
                  Clark, Spencer and
                  Wu, Elynn and
                  Perkins, Walter and
                  Kwa, Anna and
                  Duncan, James and
                  Bretherton, Christopher},
  title        = {ai2cm/ACE2.1-ERA5-AIMIP: ACE2.1-ERA5: AIMIP Phase
                   1 submission
                  },
  month        = apr,
  year         = 2026,
  publisher    = {Zenodo},
  version      = {v0.1},
  doi          = {10.5281/zenodo.19831256},
  url          = {https://doi.org/10.5281/zenodo.19831256},
}

@article{https://doi.org/10.1029/2023MS004019,
author = {Rasp, Stephan and Hoyer, Stephan and Merose, Alexander and Langmore, Ian and Battaglia, Peter and Russell, Tyler and Sanchez-Gonzalez, Alvaro and Yang, Vivian and Carver, Rob and Agrawal, Shreya and Chantry, Matthew and Ben Bouallegue, Zied and Dueben, Peter and Bromberg, Carla and Sisk, Jared and Barrington, Luke and Bell, Aaron and Sha, Fei},
title = {WeatherBench 2: A Benchmark for the Next Generation of Data-Driven Global Weather Models},
journal = {Journal of Advances in Modeling Earth Systems},
volume = {16},
number = {6},
pages = {e2023MS004019},
doi = {https://doi.org/10.1029/2023MS004019},
url = {https://agupubs.onlinelibrary.wiley.com/doi/abs/10.1029/2023MS004019},
eprint = {https://agupubs.onlinelibrary.wiley.com/doi/pdf/10.1029/2023MS004019},
note = {e2023MS004019 2023MS004019},
year = {2024}
}

@article{CINQUINI2014400,
title = {The Earth System Grid Federation: An open infrastructure for access to distributed geospatial data},
journal = {Future Generation Computer Systems},
volume = {36},
pages = {400-417},
year = {2014},
note = {Special Section: Intelligent Big Data Processing Special Section: Behavior Data Security Issues in Network Information Propagation Special Section: Energy-efficiency in Large Distributed Computing Architectures Special Section: eScience Infrastructure and Applications},
issn = {0167-739X},
doi = {https://doi.org/10.1016/j.future.2013.07.002},
url = {https://www.sciencedirect.com/science/article/pii/S0167739X13001477},
author = {Luca Cinquini and Daniel Crichton and Chris Mattmann and John Harney and Galen Shipman and Feiyi Wang and Rachana Ananthakrishnan and Neill Miller and Sebastian Denvil and Mark Morgan and Zed Pobre and Gavin M. Bell and Charles Doutriaux and Robert Drach and Dean Williams and Philip Kershaw and Stephen Pascoe and Estanislao Gonzalez and Sandro Fiore and Roland Schweitzer}
}

@Article{gmd-17-3919-2024,
AUTHOR = {Lee, J. and Gleckler, P. J. and Ahn, M.-S. and Ordonez, A. and Ullrich, P. A. and Sperber, K. R. and Taylor, K. E. and Planton, Y. Y. and Guilyardi, E. and Durack, P. and Bonfils, C. and Zelinka, M. D. and Chao, L.-W. and Dong, B. and Doutriaux, C. and Zhang, C. and Vo, T. and Boutte, J. and Wehner, M. F. and Pendergrass, A. G. and Kim, D. and Xue, Z. and Wittenberg, A. T. and Krasting, J.},
TITLE = {Systematic and objective evaluation of Earth system models: PCMDI Metrics Package (PMP) version 3},
JOURNAL = {Geoscientific Model Development},
VOLUME = {17},
YEAR = {2024},
NUMBER = {9},
PAGES = {3919--3948},
URL = {https://gmd.copernicus.org/articles/17/3919/2024/},
DOI = {10.5194/gmd-17-3919-2024}
}

@Article{gmd-9-1747-2016,
AUTHOR = {Eyring, V. and Righi, M. and Lauer, A. and Evaldsson, M. and Wenzel, S. and Jones, C. and Anav, A. and Andrews, O. and Cionni, I. and Davin, E. L. and Deser, C. and Ehbrecht, C. and Friedlingstein, P. and Gleckler, P. and Gottschaldt, K.-D. and Hagemann, S. and Juckes, M. and Kindermann, S. and Krasting, J. and Kunert, D. and Levine, R. and Loew, A. and M\"akel\"a, J. and Martin, G. and Mason, E. and Phillips, A. S. and Read, S. and Rio, C. and Roehrig, R. and Senftleben, D. and Sterl, A. and van Ulft, L. H. and Walton, J. and Wang, S. and Williams, K. D.},
TITLE = {ESMValTool (v1.0) -- a community diagnostic and performance metrics tool for
routine evaluation of Earth system models in CMIP},
JOURNAL = {Geoscientific Model Development},
VOLUME = {9},
YEAR = {2016},
NUMBER = {5},
PAGES = {1747--1802},
URL = {https://gmd.copernicus.org/articles/9/1747/2016/},
DOI = {10.5194/gmd-9-1747-2016}
}

@misc{taylor_2025_15670624,
  author       = {Taylor, Karl E. and
                  Juckes, Martin and
                  Balaji, Venkatramani and
                  Cinquini, Luca and
                  Denvil, Sebastien and
                  Durack, Paul J. and
                  Elkington, Mark and
                  Guilyardi, Eric and
                  Kharin, Slava and
                  Lautenschlager, Michael and
                  Lawrence, Bryan and
                  Nadeau, Denis and
                  Stockhause, Martina},
  title        = {CMIP6 Model Output Metadata Requirements, Data
                   Reference Syntax (DRS) and Controlled Vocabularies
                   (CVs)
                  },
  month        = jun,
  year         = 2025,
  publisher    = {Zenodo},
  version      = {6.2.8},
  doi          = {10.5281/zenodo.15670624},
  url          = {https://doi.org/10.5281/zenodo.15670624},
}

@misc{eaton_2025_17801666,
  author       = {Eaton, Brian and
                  Gregory, Jonathan and
                  Drach, Bob and
                  Taylor, Karl and
                  Hankin, Steve and
                  Caron, John and
                  Signell, Rich and
                  Bentley, Phil and
                  Rappa, Greg and
                  Höck, Heinke and
                  Pamment, Alison and
                  Juckes, Martin and
                  Raspaud, Martin and
                  Blower, Jon and
                  Horne, Randy and
                  Whiteaker, Timothy and
                  Blodgett, David and
                  Zender, Charlie and
                  Lee, Daniel and
                  Hassell, David and
                  Snow, Alan D. and
                  Kölling, Tobias and
                  Allured, Dave and
                  Jelenak, Aleksandar and
                  Soerensen, Anders Meier and
                  Gaultier, Lucile and
                  Herlédan, Sylvain and
                  Manzano, Fernando and
                  Bärring, Lars and
                  Barker, Christopher and
                  Bartholomew, Sadie L. and
                  Lavergne, Thomas and
                  Lawrence, Bryan and
                  Massey, Neil and
                  Cofiño, Antonio S. and
                  McGinnis, Seth and
                  Laake, Patrick Van},
  title        = {NetCDF Climate and Forecast (CF) Metadata
                   Conventions
                  },
  month        = dec,
  year         = 2025,
  publisher    = {CF Community},
  version      = {1.13},
  doi          = {10.5281/zenodo.17801666},
  url          = {https://doi.org/10.5281/zenodo.17801666},
}

@misc{henn_2026_21302180,
  author       = {Henn, Brian and
                  Bretherton, Christopher and
                  Koldunov, Nikolay V. and
                  Watt-Meyer, Oliver},
  title        = {ai2cm/AIMIP: Manuscript preprint release, v0.3},
  month        = jul,
  year         = 2026,
  publisher    = {Zenodo},
  version      = {v0.3},
  doi          = {10.5281/zenodo.21302180},
  url          = {https://doi.org/10.5281/zenodo.21302180},
}

@misc{liu2022convnet2020s,
      title={A ConvNet for the 2020s}, 
      author={Zhuang Liu and Hanzi Mao and Chao-Yuan Wu and Christoph Feichtenhofer and Trevor Darrell and Saining Xie},
      year={2022},
      eprint={2201.03545},
      archivePrefix={arXiv},
      primaryClass={cs.CV},
      url={https://arxiv.org/abs/2201.03545}, 
}

@article{Karlbauer2024,
author = {Karlbauer, Matthias and Cresswell-Clay, Nathaniel and Durran, Dale R. and Moreno, Raul A. and Kurth, Thorsten and Bonev, Boris and Brenowitz, Noah and Butz, Martin V.},
title = {Advancing Parsimonious Deep Learning Weather Prediction Using the HEALPix Mesh},
journal = {Journal of Advances in Modeling Earth Systems},
volume = {16},
number = {8},
pages = {e2023MS004021},
doi = {https://doi.org/10.1029/2023MS004021},
url = {https://agupubs.onlinelibrary.wiley.com/doi/abs/10.1029/2023MS004021},
eprint = {https://agupubs.onlinelibrary.wiley.com/doi/pdf/10.1029/2023MS004021},
note = {e2023MS004021 2023MS004021},
year = {2024}
}

@misc{guillaume_couairon_2026_20784771,
  author       = {Guillaume Couairon and
                  Renu Singh and
                  Brunstein, Robert and
                  Cousin, Raphael and
                  Hinderer, Sébastien and
                  Adrien Le Coz and
                  Díaz, Mauricio},
  title        = {INRIA/geoarches: v0.2.0 (Dev): AIMIP adaptation
                   Pre-release
                  },
  month        = jun,
  year         = 2026,
  publisher    = {Zenodo},
  version      = {v0.2.0-dev},
  doi          = {10.5281/zenodo.20784771},
  url          = {https://doi.org/10.5281/zenodo.20784771},
}

@misc{manshausen_2026_20832634,
  author       = {Manshausen, Peter and
                  Brenowitz, Noah and
                  Berner, Julius and
                  Kashinath, Karthik and
                  Pritchard, Michael},
  title        = {Checkpoints and code for: Towards accurate extreme
                   event likelihoods from diffusion model climate
                   emulators
                  },
  month        = jun,
  year         = 2026,
  publisher    = {Zenodo},
  doi          = {10.5281/zenodo.20832634},
  url          = {https://doi.org/10.5281/zenodo.20832634},
}

@misc{nathanielcresswellclay_2026_21270137,
  author       = {Cresswell-Clay, Nathaniel},
  title        = {nathanielcresswellclay/DLESyM\_aimip-2026:
                   AIMIP-2026
                  },
  month        = jul,
  year         = 2026,
  publisher    = {Zenodo},
  version      = {v1.0},
  doi          = {10.5281/zenodo.21270137},
  url          = {https://doi.org/10.5281/zenodo.21270137},
}

@misc{kyle_hall_2026_21430292,
  author       = {Kyle Hall and
                  Maria J. Molina},
  title        = {kjhall01/monthly-diffusion: MDv0.9 AIMIP Release
                   (Licensed)
                  },
  month        = jul,
  year         = 2026,
  publisher    = {Zenodo},
  version      = {MDv0.9-1.5-licensed},
  doi          = {10.5281/zenodo.21430292},
  url          = {https://doi.org/10.5281/zenodo.21430292},
}

@misc{kochkov_2026_21303721,
  author       = {Kochkov, Dmitrii and
                  Hoyer, Stephan and
                  Yuval, Janni},
  title        = {Terrax code (WIP)},
  month        = jul,
  year         = 2026,
  publisher    = {Zenodo},
  doi          = {10.5281/zenodo.21303721},
  url          = {https://doi.org/10.5281/zenodo.21303721},
}

@article{https://doi.org/10.1029/2019MS001829,
author = {Held, I. M. and Guo, H. and Adcroft, A. and Dunne, J. P. and Horowitz, L. W. and Krasting, J. and Shevliakova, E. and Winton, M. and Zhao, M. and Bushuk, M. and Wittenberg, A. T. and Wyman, B. and Xiang, B. and Zhang, R. and Anderson, W. and Balaji, V. and Donner, L. and Dunne, K. and Durachta, J. and Gauthier, P. P. G. and Ginoux, P. and Golaz, J.-C. and Griffies, S. M. and Hallberg, R. and Harris, L. and Harrison, M. and Hurlin, W. and John, J. and Lin, P. and Lin, S.-J. and Malyshev, S. and Menzel, R. and Milly, P. C. D. and Ming, Y. and Naik, V. and Paynter, D. and Paulot, F. and Ramaswamy, V. and Reichl, B. and Robinson, T. and Rosati, A. and Seman, C. and Silvers, L. G. and Underwood, S. and Zadeh, N.},
title = {Structure and Performance of GFDL's CM4.0 Climate Model},
journal = {Journal of Advances in Modeling Earth Systems},
volume = {11},
number = {11},
pages = {3691-3727},
doi = {https://doi.org/10.1029/2019MS001829},
url = {https://agupubs.onlinelibrary.wiley.com/doi/abs/10.1029/2019MS001829},
eprint = {https://agupubs.onlinelibrary.wiley.com/doi/pdf/10.1029/2019MS001829},
year = {2019}
}

@misc{singh2026evaluatingskillstabilityarchesweather,
      title={Evaluating Skill and Stability of ArchesWeather and ArchesWeatherGen under Multi-Decadal Climate Simulations}, 
      author={Renu Singh and Robert Brunstein and Antonia Jost and Yana Hasson and Thomas Rackow and Claire Monteleoni and Christian Lessig and Guillaume Couairon},
      year={2026},
      eprint={2605.29976},
      archivePrefix={arXiv},
      primaryClass={physics.ao-ph},
      url={https://arxiv.org/abs/2605.29976}, 
}

@misc{era5_hourly_single_levels,
  author       = {Hersbach, H. and Bell, B. and Berrisford, P. and Biavati, G. and Hor{\'a}nyi, A. and Mu{\~n}oz Sabater, J. and Nicolas, J. and Peubey, C. and Radu, R. and Rozum, I. and Schepers, D. and Simmons, A. and Soci, C. and Dee, D. and Th{\'e}paut, J.-N.},
  title        = {{ERA5 hourly data on single levels from 1940 to present}},
  year         = {2018},
  howpublished = {Copernicus Climate Change Service (C3S) Climate Data Store (CDS)},
  doi          = {10.24381/cds.adbb2d47},
  note         = {Accessed: 23 July 2026}
}

@misc{era5_hourly_pressure_levels,
  author       = {Hersbach, H. and Bell, B. and Berrisford, P. and Biavati, G. and Hor{\'a}nyi, A. and Mu{\~n}oz Sabater, J. and Nicolas, J. and Peubey, C. and Radu, R. and Rozum, I. and Schepers, D. and Simmons, A. and Soci, C. and Dee, D. and Th{\'e}paut, J.-N.},
  title        = {{ERA5 hourly data on pressure levels from 1940 to present}},
  year         = {2018},
  howpublished = {Copernicus Climate Change Service (C3S) Climate Data Store (CDS)},
  doi          = {10.24381/cds.bd0915c6},
  note         = {Accessed: 23 July 2026}
}

@misc{https://doi.org/10.22033/ESGF/CMIP6.8508,
      url = {https://doi.org/10.22033/ESGF/CMIP6.8508},
      title = {NOAA-GFDL GFDL-CM4 model output prepared for CMIP6 CFMIP amip-p4K},
      publisher = {Earth System Grid Federation},
      year = {2018},
      author = {Silvers, Levi and Blanton, Chris and McHugh, Colleen and John, Jasmin G and Radhakrishnan, Aparna and Rand, Kristopher and Balaji, V and Dupuis, Christopher and Durachta, Jeff and Guo, Huan and Hemler, Richard and Lin, Pu and Nikonov, Serguei and Paynter, David J and Ploshay, Jeffrey and Vahlenkamp, Hans and Wilson, Chandin and Wyman, Bruce and Robinson, Thomas and Zeng, Yujin and Zhao, Ming},
      doi = {10.22033/ESGF/CMIP6.8508}
}


\newpage

\appendix

\section{AIMIP Phase 1 monthly SST and SIC dataset}
\label{appendix:aimip_forcing}


The CMIP input4mips project that assembles the needed forcing data for the DECK simulations provides AMIP specifications of the monthly historical SST and SIC.  However, it is not suitable for AIMIP Phase 1.  First, it does not extend past 2022, while AIMIP Phase 1 inference simulations cover through 2024 to maximize the possible length of high-quality observational comparison.  Second, the AMIP algorithm for calculating monthly values for SST and SIC is problematic.  It involves specifying mid-month values that, when linearly interpolated in time, give the monthly-mean values in the reference dataset.  This inevitably produces overshoots in the mid-month values.  SIC in some grid cells can switch between near 1 and near 0 in successive months, and the CMIP algorithm occasionally results in mid-month values of SIC that are below zero and must be thresholded to zero. This results in small biases in annual-mean SIC that have a noticeable effect on the annual mean temperature in some grid cells in the seasonal ice zone.

Instead, we created a compact 1979-2024 monthly AMIP-like SST and SIC forcing dataset to use for AIMIP Phase 1 inference runs that addresses these issues \citep{arcomano_2025_17065758}. It is based on daily outputs from ERA5 on its 0.25° latitude-longitude grid.  These are averaged to forcing values at the beginning of each month using a centered rectangular averaging window between the midpoints of the previous and current months (to enable linear interpolation during Dec. 2024, the dataset extends until Jan. 1 2025).  The SST and SIC forcings at intermediate times are obtained by linear interpolation.  This method of generating monthly forcings doesn’t produce data overshoots and preserves the annual time-mean of each forcing field, although individual monthly means are not exactly preserved.  We have checked that when used to force inference runs with the ACE2.1-ERA5 emulator, it produces a climate nearly identical to the use of daily forcing data, even in the seasonal sea ice zones.  Each modeling group may spatially interpolate this monthly forcing to their native grid, e.g. using the conservative regridding option of xesmf.

This ERA5-based SIC forcing data has small incompatibilities between the land mask and the SIC within 0.25° cells on polar coastlines and lake boundaries.  This is native to ERA5, and is even present in the ERA5 data on a reduced Gaussian grid.  The easiest work-around is to limit SIC to 1 - land fraction.  It might be better to fill the resulting ‘lost’ sea ice into the closest adjacent coastal cells that have SIC + land fraction < 1, but the small added benefit is probably not worth the required effort. 

\section{cBottle1.3 physics indices checkpoints}  
\label{appendix:cBottle_indices}
The following describes the checkpoints used to generate the cBottle `physics' realizations, which generate its ensemble of simulations in AIMIP Phase 1. cBottle1.3, like the published version, is an Ensemble-of-Experts model. Different parts of the denoising are carried out by different networks, with the higher noise levels being denoised by less trained/early-stopped versions of the network. This is to avoid overfitting at large noise levels (see \cite{brenowitz2025climatebottlegenerativefoundation} for details). For every model, we use three networks, the first active at noise lower than 10, the second between 10 and 100, and the third over 100. Numbers indicate the amount of noisy samples this network is trained on.

\textbf{Physics Indices}:

\begin{itemize}
    \item p1 checkpoints: 
    \begin{itemize}
        \item training-state-000512000.checkpoint
        \item training-state-002048000.checkpoint
        \item training-state-009856000.checkpoint
    \end{itemize}

    \item p2 checkpoints: 
    \begin{itemize}
        \item  training-state-000512000.checkpoint
        \item  training-state-002176000.checkpoint
        \item  training-state-009984000.checkpoint
    \end{itemize}

    \item p3 checkpoints: 
    \begin{itemize} 
        \item  training-state-000640000.checkpoint
        \item  training-state-002048000.checkpoint
        \item  training-state-010112000.checkpoint
    \end{itemize}

    \item p4 checkpoints: 
    \begin{itemize} 
        \item  training-state-000640000.checkpoint
        \item  training-state-002176000.checkpoint
        \item  training-state-009728000.checkpoint
    \end{itemize}

    \item p5: Same checkpoints as p4, but latent space is uncorrelated in time, so every sample is fully independent.

\end{itemize}

\section{Additional evaluation results}

\subsection{Biases}
\label{additional_biases}

\begin{figure*}[h]
\includegraphics[width=17.4cm]{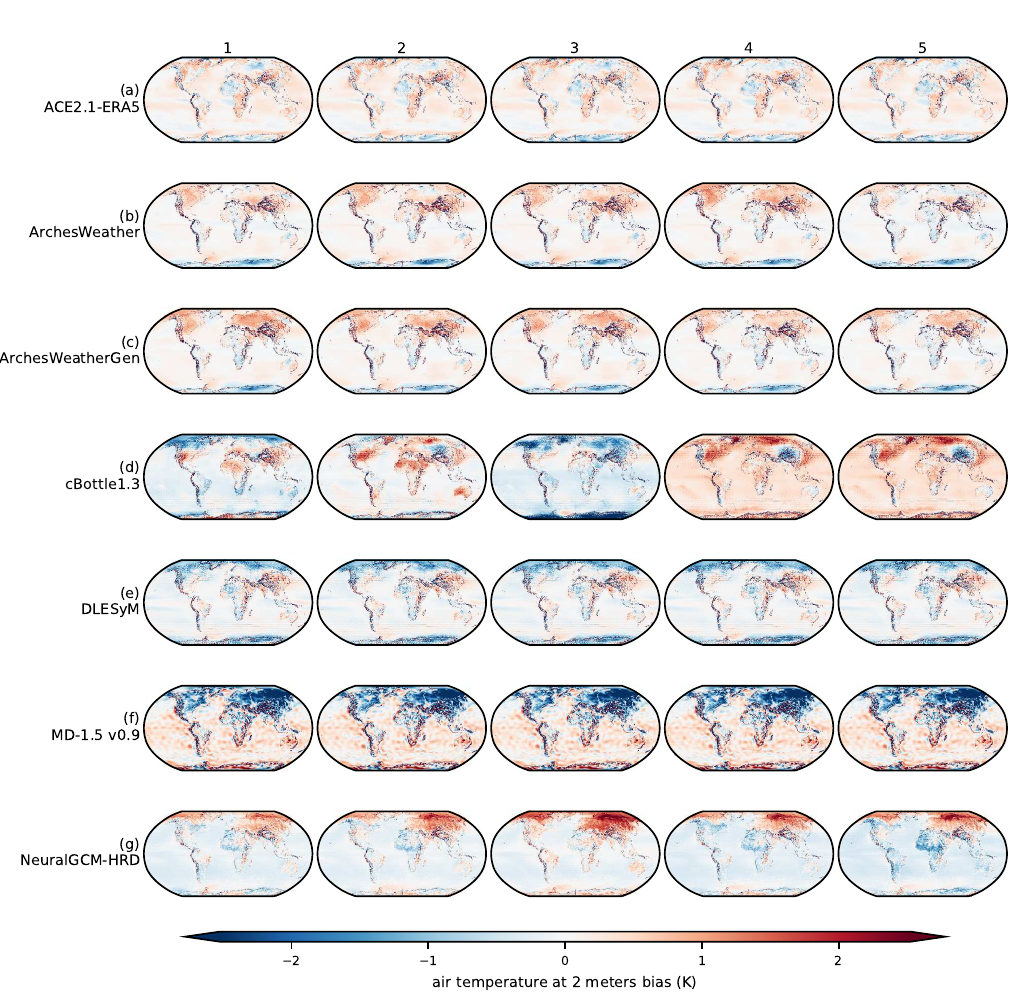}
\caption{Bias of 2-meter air temperature over the training period (1979-2014) for each model ensemble member.}
\label{app_fig:bias_map_train_tas_ensemble}
\end{figure*}

\begin{figure*}[h]
\includegraphics[width=17.4cm]{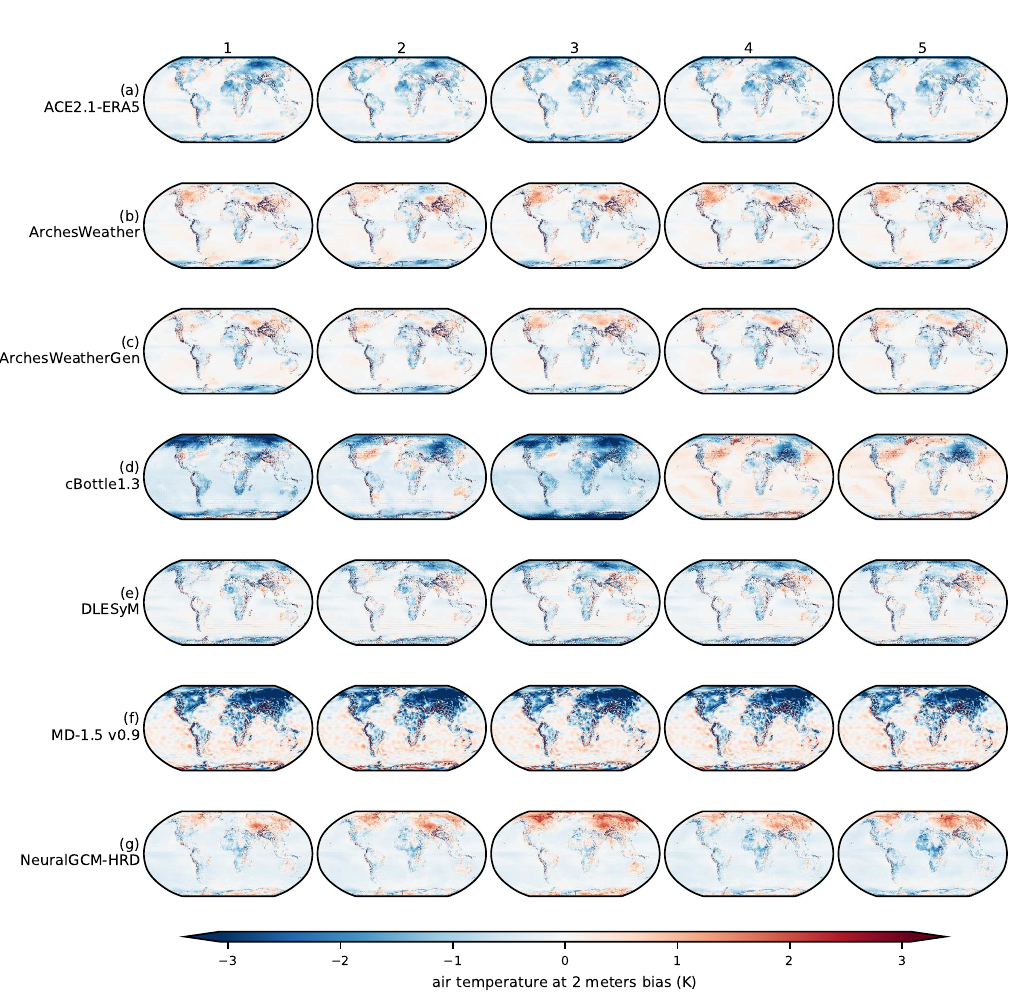}
\caption{Bias of 2-meter air temperature over the test period (2015-2024) for each model ensemble member.}
\label{app_fig:bias_map_test_tas_ensemble}
\end{figure*}

\begin{figure*}[h]
\includegraphics[width=17.4cm]{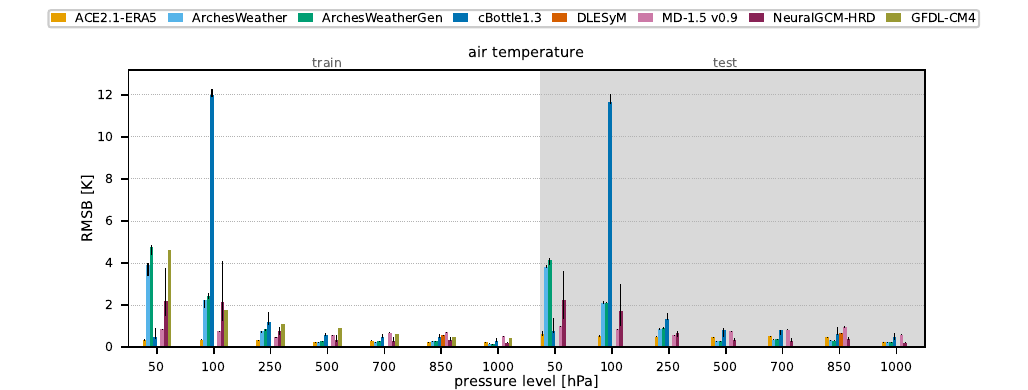}
\caption{RSMB for air temperature over pressure levels and training and test periods.}
\label{app_fig:rmsb_ta_plev}
\end{figure*}

\begin{figure*}[h]
\includegraphics[width=17.4cm]{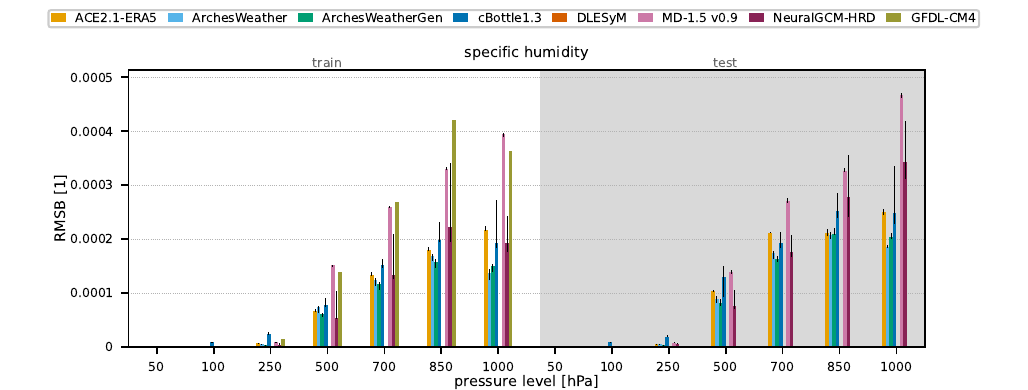}
\caption{RSMB for specific humidity over pressure levels and training and test periods.}
\label{app_fig:rmsb_hus_plev}
\end{figure*}

\begin{figure*}[h]
\includegraphics[width=17.4cm]{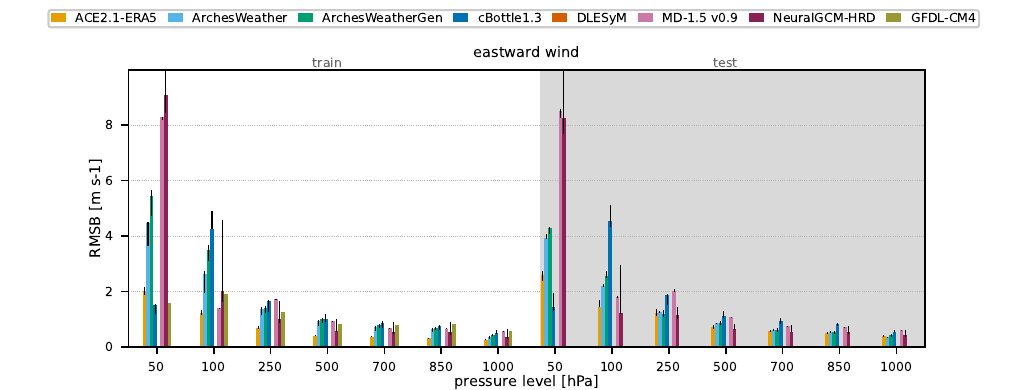}
\caption{RSMB for eastward wind over pressure levels and training and test periods.}
\label{app_fig:rmsb_ua_plev}
\end{figure*}

\begin{figure*}[h]
\includegraphics[width=17.4cm]{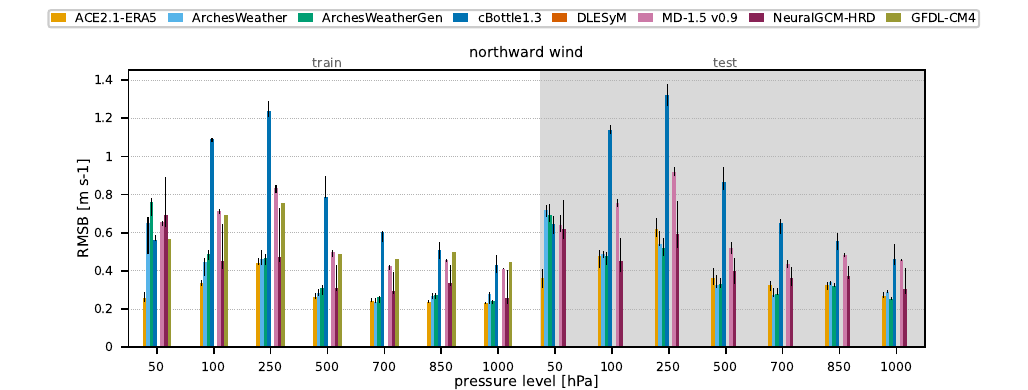}
\caption{RSMB for northward wind over pressure levels and training and test periods.}
\label{app_fig:rmsb_va_plev}
\end{figure*}

In Figs. \ref{app_fig:bias_map_train_tas_ensemble} and \ref{app_fig:bias_map_test_tas_ensemble}, we show the 2-meter air temperature biases over the training and test periods for each model ensemble. In Figs. \ref{app_fig:rmsb_ta_plev}, \ref{app_fig:rmsb_hus_plev}, \ref{app_fig:rmsb_ua_plev}, and \ref{app_fig:rmsb_va_plev}, we show RMSB at 1\degree ~resolution over 7 pressure levels for air temperature, specific humidity, and eastward and northward wind, respectively.

\subsection{Trends}
\label{additional_trends}

\begin{figure*}[h]
\includegraphics[width=17.4cm]{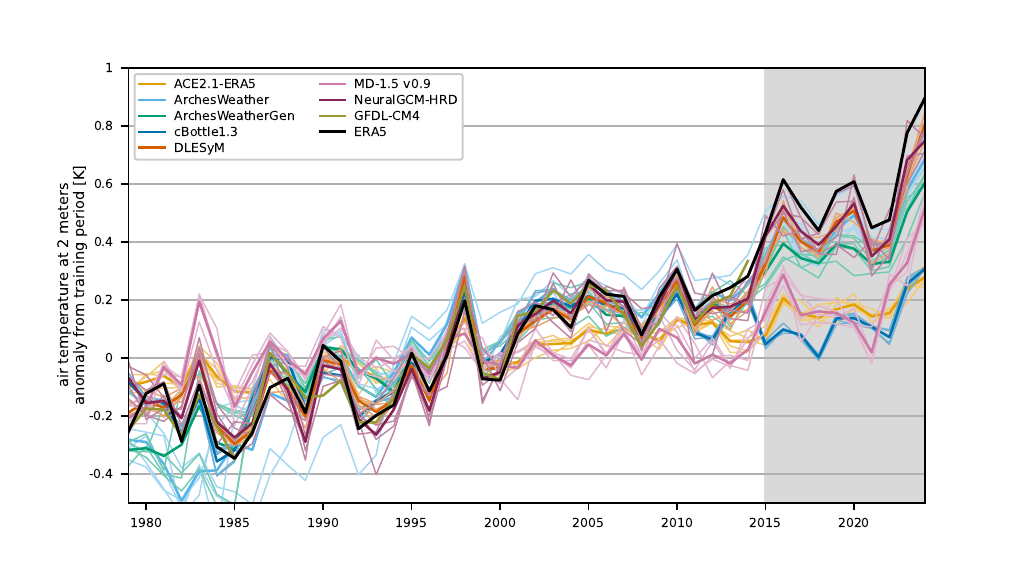}
\caption{Annual- and global-mean 2-meter air temperature series as in Fig. \ref{fig:trend_tas}, but including the models' ensemble members (thin lines).}
\label{app_fig:trend_tas_ens}
\end{figure*}

\begin{figure*}[h]
\includegraphics[width=17.4cm]{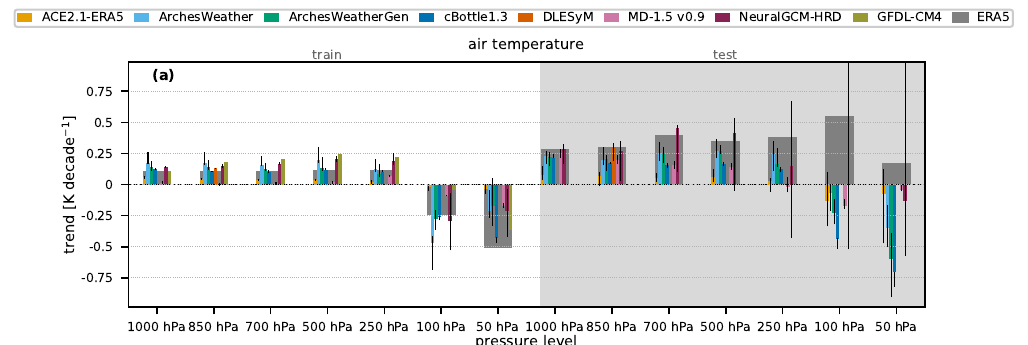}
\caption{Global mean trends at 1\degree ~resolution as in Fig. \ref{fig:trend_bar}, but for air temperature over pressure levels and training and test periods.}
\label{app_fig:trend_bar_ta_plev}
\end{figure*}

\begin{figure*}[h]
\includegraphics[width=17.4cm]{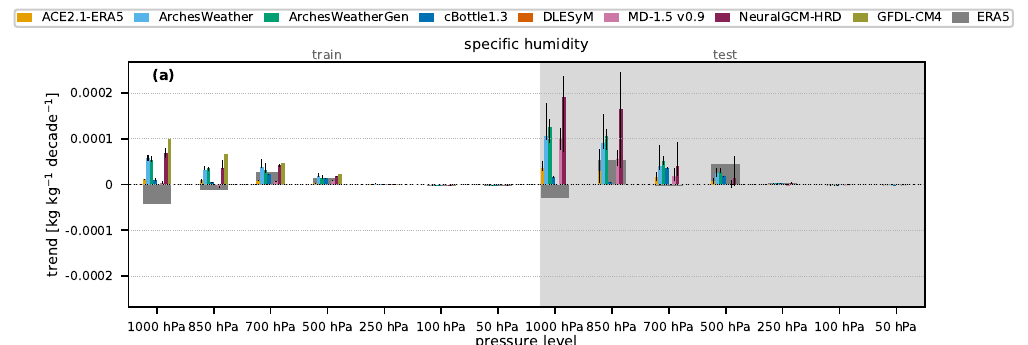}
\caption{Global mean trends at 1\degree ~resolution as in Fig. \ref{fig:trend_bar}, but for specific humidity over pressure levels and training and test periods.}
\label{app_fig:trend_bar_hus_plev}
\end{figure*}

In Fig. \ref{app_fig:trend_tas_ens}, we show the annual- and global-mean 2-meter air temperature series for each AIWCM, including its ensemble members. Figs. \ref{app_fig:trend_bar_ta_plev} and \ref{app_fig:trend_bar_hus_plev} show trends in air temperature and specific humidity, respectively, over pressure levels at 1\degree ~resolution.

\subsection{ENSO response}

\begin{figure*}[h]
\includegraphics[width=17.4cm]{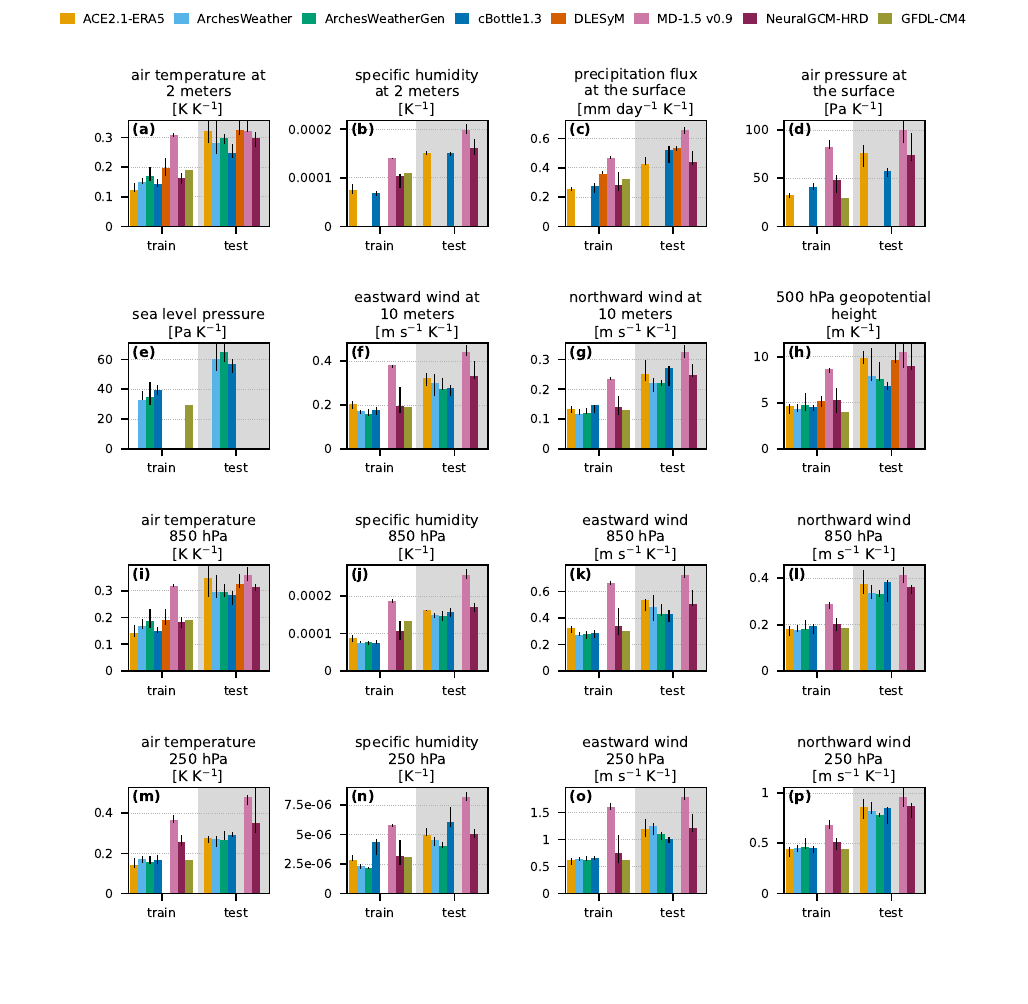}
\caption{Root-mean-square error (RMSE) of ENSO coefficients versus ERA5 coefficients, area-weighted over the globe on the 1\degree ~grid, for the same variable set as Fig. \ref{fig:rmsb_summary}.}
\label{app_fig:enso_rmse}
\end{figure*}

Figure \ref{app_fig:enso_rmse} shows globally-averaged ENSO coefficient errors.

\subsection{Daily variability}
\label{additional_daily_variability}

\begin{figure*}[h]
\includegraphics[width=17.4cm]{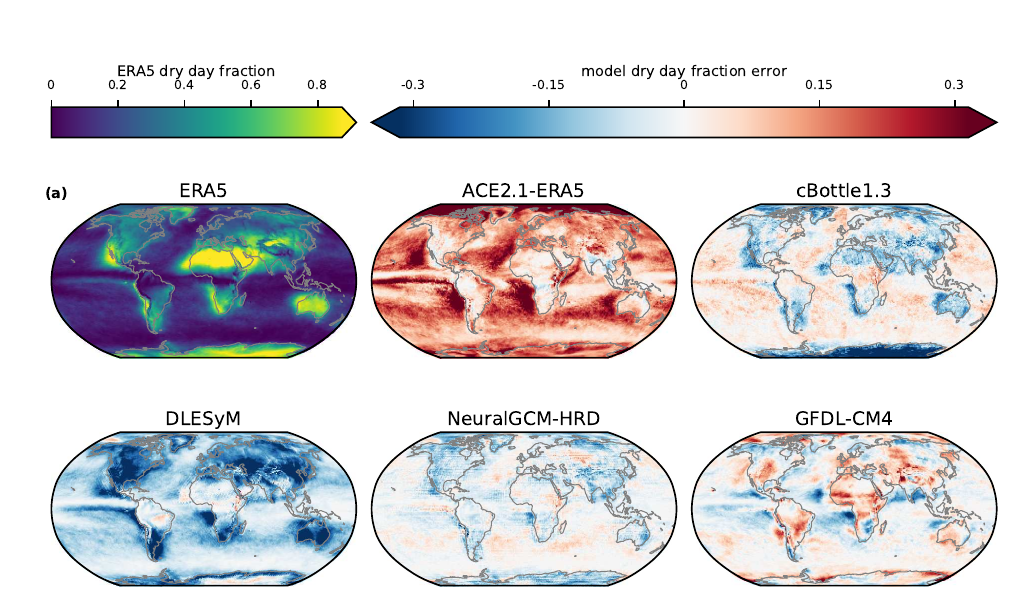}
\caption{Dry-day fraction error in ERA5 (top left panel) and dry day fraction errors versus ERA5 (subsequent panels). Computation is over 1979 and a cutoff of 0.1 mm is used to define a dry day.}
\label{app_fig:dry_day_frac}
\end{figure*}

Figure \ref{app_fig:dry_day_frac} shows dry-day fraction errors versus ERA5 over 1979 at 1\degree resolution, for models that submitted daily surface precipitation.

\clearpage

\section{Selected results at 2.8\degree ~resolution}
\label{app_2p8deg_results}

\begin{figure*}[h]
\includegraphics[width=17.4cm]{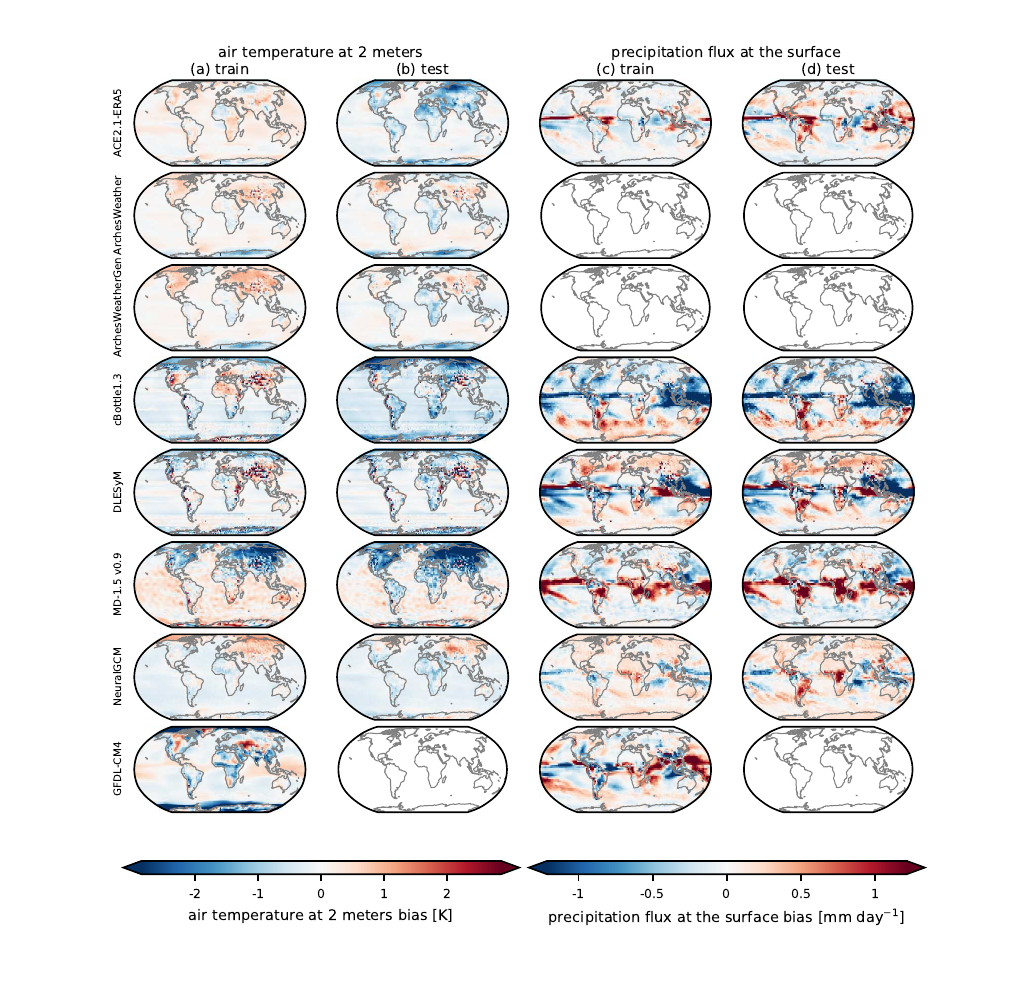}
\caption{Biases as in Fig. \ref{fig:bias_map}, but at 2.8\degree ~resolution.}
\label{app_fig:bias_map_2p8deg}
\end{figure*}

\begin{figure*}[h]
\includegraphics[width=17.4cm]{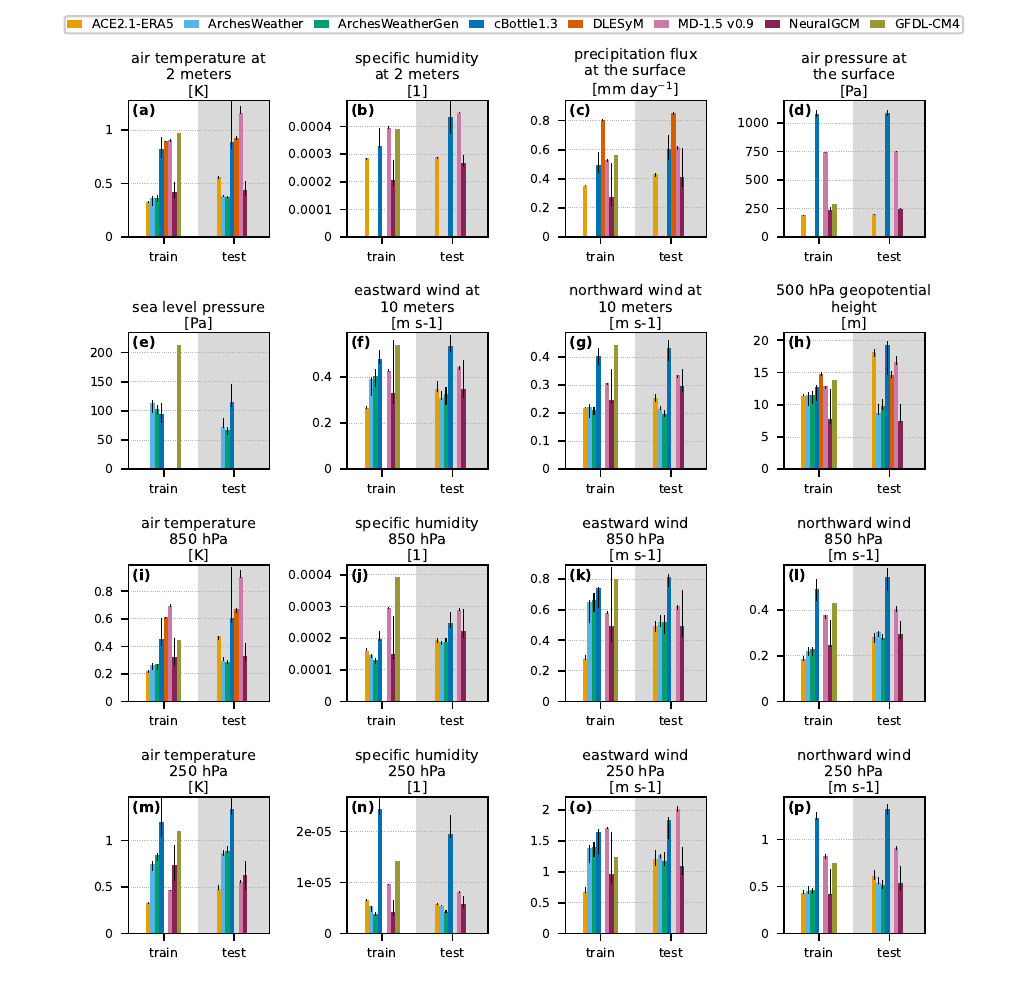}
\caption{RSMB as in Fig. \ref{fig:rmsb_summary}, but at 2.8\degree ~resolution.}
\label{app_fig:rsmb_submmary_2p8deg}
\end{figure*}

\begin{figure*}[h]
\includegraphics[width=17.4cm]{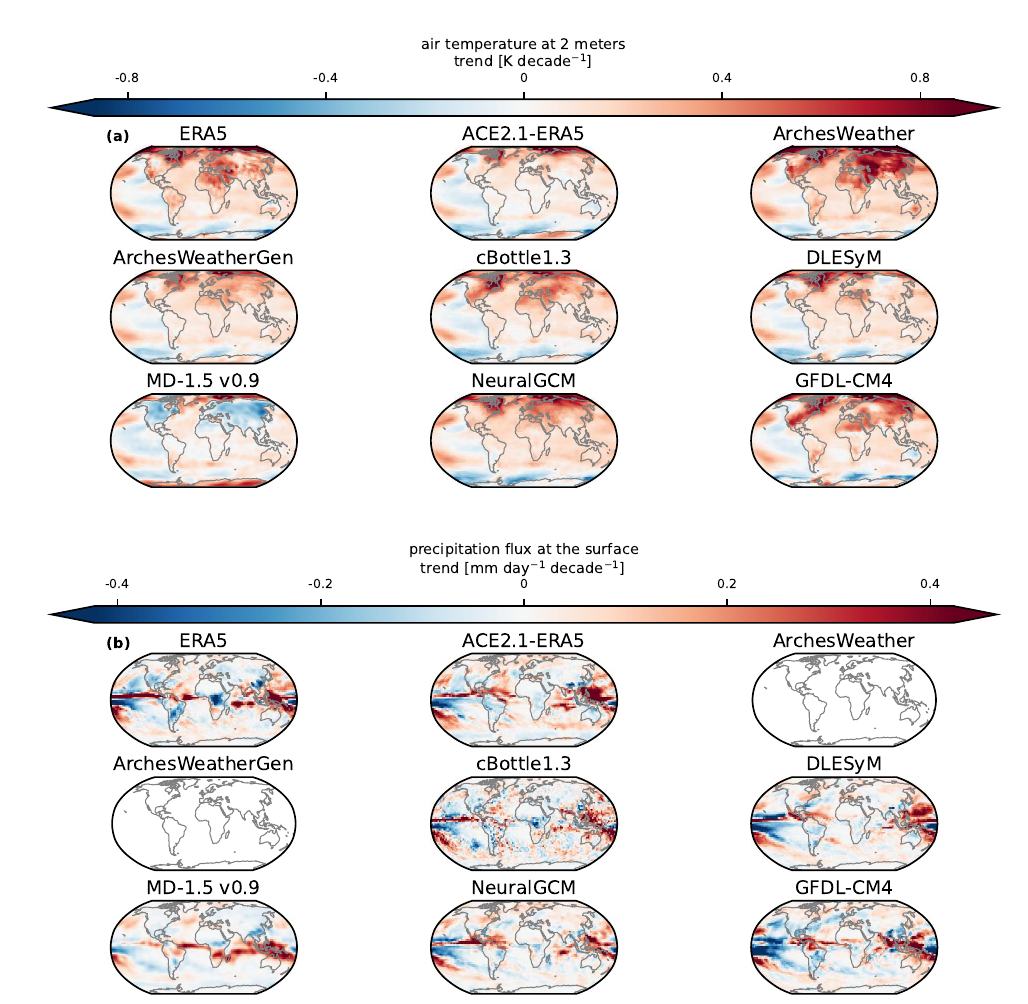}
\caption{Global-mean trends as in Fig. \ref{fig:trend_maps}, but at 2.8\degree ~resolution}
\label{app_fig:trend_maps_2p8deg}
\end{figure*}

\begin{figure*}[h]
\includegraphics[width=17.4cm]{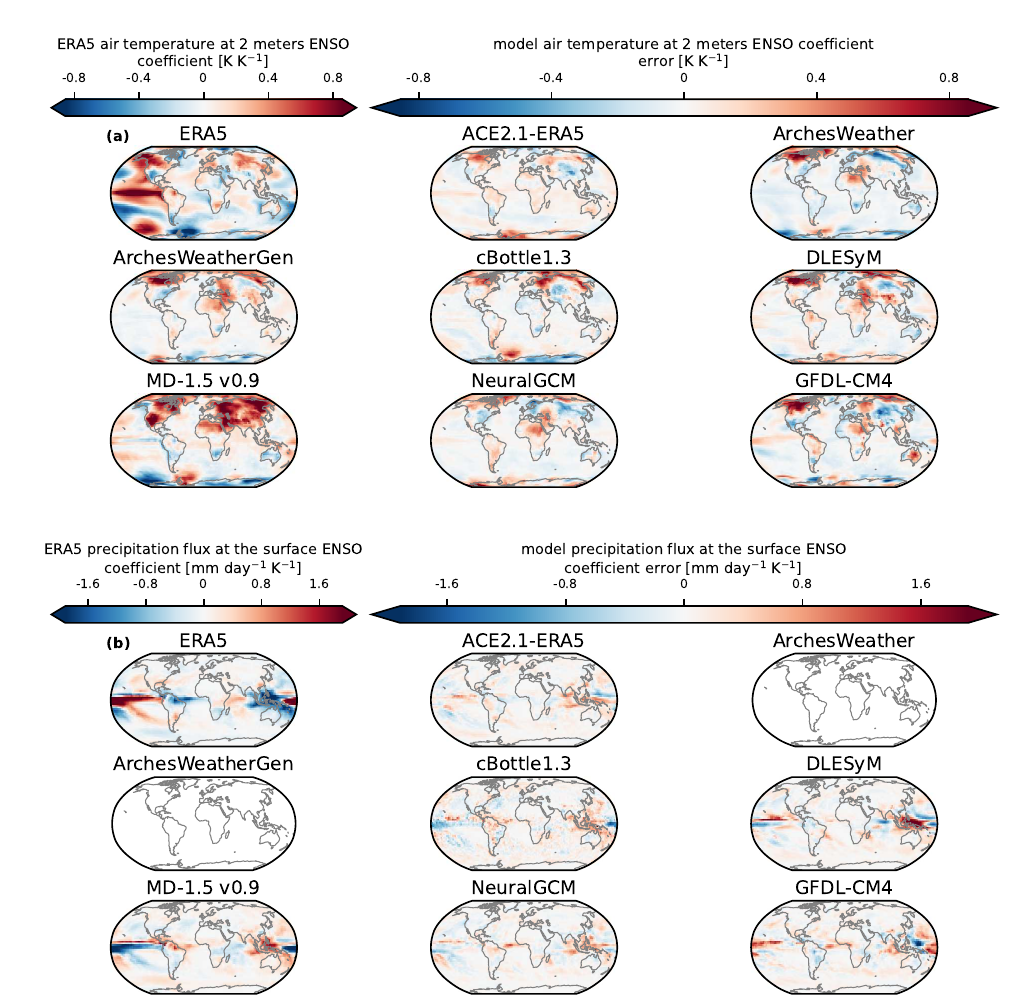}
\caption{ERA5 ENSO coefficients and model coefficient errors as in Fig. \ref{fig:enso_maps}, but at 2.8\degree ~resolution.}
\label{app_fig:enso_maps_2p8deg}
\end{figure*}

\begin{figure*}[h]
\includegraphics[width=17.4cm]{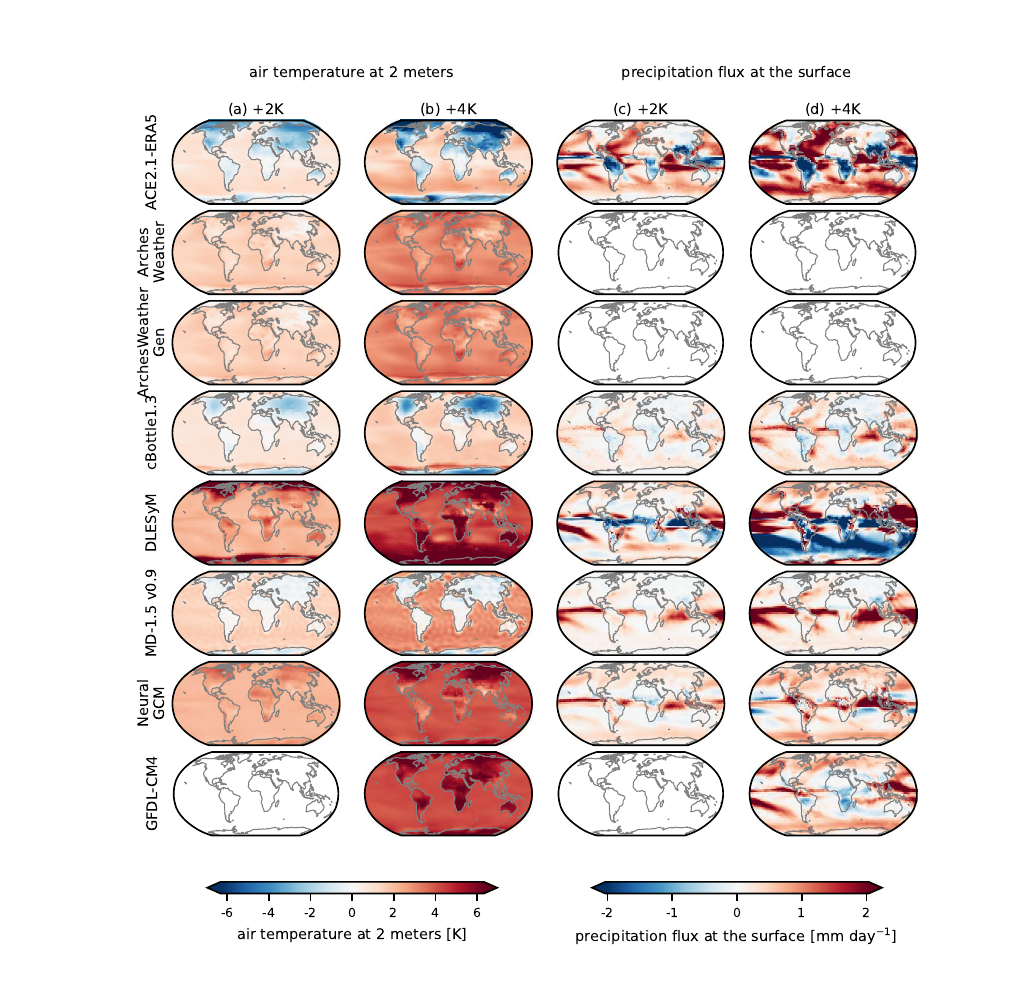}
\caption{As in Fig. \ref{fig:perturbation_response_map}, but at 2.8\degree ~resolution.}
\label{app_fig:perturbation_response_map_2p8deg}
\end{figure*}

We show selected results at 2.8\degree ~resolution, with NeuralGCM instead of NeuralGCM-HRD. In Figs. \ref{app_fig:bias_map_2p8deg} and \ref{app_fig:rsmb_submmary_2p8deg}, we show bias maps and RMSB similar to Figs. \ref{fig:bias_map} and \ref{fig:rmsb_summary}, but at 2.8\degree ~resolution. In Fig. \ref{app_fig:trend_maps_2p8deg}, we show trend maps similar to Fig. \ref{fig:trend_maps}, but at 2.8\degree ~resolution. Figure \ref{app_fig:enso_maps_2p8deg} shows ENSO coefficient errors as in Fig. \ref{fig:enso_maps}, but at 2.8\degree ~resolution. In Fig. \ref{app_fig:perturbation_response_map_2p8deg}, we show perturbation responses similar to Fig. \ref{fig:perturbation_response_map}, but at 2.8\degree ~resolution.

\noappendix       

\end{document}